\def\R{{\rm I}\!{\rm R}}

\def\carre{\hbox{\vrule \vbox to 7pt{\hrule width 6pt \vfill \hrule}\vrule }}
\parindent = 0cm

\font\un = cmbx10 at 14pt

\centerline{\un The  semi-classical  limit of the  time dependent}
\medskip

\centerline {\un   Hartree-Fock  equation.}

\bigskip \centerline {\un I.  The Weyl  Symbol  of the  solution.}

\bigskip

 \centerline {\bf L. Amour, M. Khodja et J. Nourrigat}

 \bigskip

 \centerline {Laboratoire de Math\'ematiques,  EA4535}

\medskip

 \centerline {Universit\'e de Reims, U.F.R. Sciences Exactes et Naturelles}

\medskip

 \centerline {Moulin de la Housse - BP 1039, 51687 REIMS Cedex 2, France}
\medskip

 \centerline {and FR.CNRS.3399}

\vskip 1cm


{\bf  1.   Introduction.}

\bigskip

   The   time  dependent  Hartree-Fock  (TDHF)
   describes, when  the number  $N$  of  particules tends
   to  infinity,  the time evolution  of the density operator of a quantum system
  of  $N$  interacting particules  in the mean field approximation, in other words,
  when the interaction
   between  two  particules  is of order  $1/N$.  For this limiting process  when
   $N\rightarrow + \infty$,  see, for eaxample,  the recent works  of
   [AN1], [AN2], [BGGM] , [ES], [FKS],....

   \bigskip

   A  solution to this  equation  (TDHF) is a density operator
   $\rho_h(t)$, that is to say   a trace class  operator  in  ${\cal H} =
   L^2(\R^n)$  (if each of the  particules in interaction  is moving
   in  $\R^n$),  selfadjoint, $\geq 0$, of trace equal to  1,
   evolving  as a function of $t$,  and depending  on  a
    semiclassical parameter  $h$   tending  to 0.

   \bigskip

   The  aim of this work  is the analysis  of such a  solution when the parameter
   $h$ tends  to $0$. The analysis  will, in particular,
   make  precise  the relationship  to the  Vlasov  equation. To establish  this relationship,
   we may associate  to the operator  $\rho_h(t)$  two  notions of
   symbols:  the Weyl  symbol  (studied in  this paper), and  the
   Wick  symbol  (studied  in a second paper). The
    Weyl symbol  may only  be  used under rather strong  hypotheses
    on the operator $\rho_h(0)$, hypotheses that will be weakened
    in the second part. In the two
    articles, the convergence of different symbols  when
    $h \rightarrow 0$  will be a  convergence in  $L^1(\R^{2n})$.

\bigskip

    Since  (TDHF) appears as  a  limiting process   when the number of particules
    $N$ de particules tend  to infinity,   a natural  question  is the one  of the interchange
    of the two limits,   the one  where $N$ tends  to infinity, and the one where
    the  semiclassical  parameter $h$ tends vers $0$. It  is the subject of the
     article [PP]  of  F. Pezzotti and  M. Pulvirenti,  where it is shown that
     the  Weyl symbol  of the marginal  density operator
     associated  to a  particule in a  system of $N$
     particules,
     admits   an   asymptotic expansion in  powers of  $h$, and  that
     the Weyl symbol  tends   towards   the symbol  of a solution of  (TDHF), and
     that the  coefficient of  $h^j$ in the asymptotic expansion  of the symbol
     admits a limit when  $N$ tends  to  infinity , and that
      for  $j=0$, this  limit is a  solution of the Vlasov   equation.
     We observe  that in  [PP]  les limits are in the sense of
     ${\cal S}'(\R^{2n})$, while in this work  and in the second
     part   they are in the sense of $L^1(\R^{2n})$. See also [P]
     and [GMP].

\bigskip

The  two  equations (TDHF and  Vlasov) describe the time evolution
of  particle density,  the first one in quantum mechanics, and the
second one in classical mechanics. Both are dependent on two
potentials $V$ et $W$.  The first one, $V$, is the external
potential to which all particles are submitted, while
 $W$ is the mutual interaction potential. In this work,
these two functions $V$ and  $W$ will be  real valued  $C^{\infty}$
functions on $\R^n$, assumed to be  bounded  as well as all their
derivatives. Let us state precisely now these two equations.

\bigskip

In classical mechanics, the density at time  $t$ is a function $v(x
, \xi , t)\geq 0$  in  $L^1 (\R^{2n})$, whose integral  equals  $1$.
The classical  average potential at the point $x$  associated  to
the  density $v(., t)$ is
$$ V_{cl} (x, v(., t) )   = V(x) +  \int _{\R^{2n}} W(x -y) v(y , \eta , t)
dy d\eta .  \leqno (1.1) $$
The  function $v$ verifies  the  Vlasov  equation:
$$ {\partial v \over \partial t }  + 2 \sum _{j=1}^n \xi_j {\partial
v \over \partial x_j}  -  \sum _{j=1}^n  {\partial V_{cl }(x , v(.,
t) ) \over \partial x_j} {\partial v \over \partial \xi _j} =0.
\leqno (1.2) $$
If,  at an initial  time, the  data $v(. , 0)$ is in  $L^1
(\R^{2n})$, and if it is  $\geq 0$, these two  properties  remain
true  for all  $t\in \R$,  and the  integral $\R^{2n}$ de $v(., t)$
is conserved. (see  Braun-Hepp [BH]).

\bigskip

In quantum mechanics, the  density  at time  $t$  is described by  a
family of self adjoint  trace class operators    $\rho_h (t)\geq 0$
($h>0$) on ${\cal H} = L^2(\R^n)$, with trace equals   to $1$.  The
average quantum  potential   at the point  $x$, associated    to the
density operator  $\rho_h(t)$, is:
$$ V_{q} (x , \rho_h(t) ) =  V(x) + Tr ( W_x \rho_h (t )), \leqno (1.3)$$
where  $W_x$ is   the  multiplication operator by  the  function $y
\rightarrow W(x -y)$.  We will denote by  $V_{q} (\rho_h (t))$ the
multiplication operator  by  the function $x\rightarrow V_{q} (x ,
\rho_h(t) )$. The  density  operator  $\rho_h(t)$ is  trace class,
self adjoint, $\geq 0$, and verifies  the (TDHF) equation :
$$ ih {\partial \over \partial t} \rho_h (t) = - {h^2 }
[ \Delta , \rho_h(t ) ] + [ V_{q} (\rho_h(t)) , \rho_h (t )], \leqno
(1.4)$$
where  $\Delta $ is the Laplacian. To make precise the meaning given
to the commutators, let us  recall the notion of classical solution
(TDHF) introduced  by Bove da Prato Fano ([BdPF1], [BdPF2]). Let us
denote by ${\cal L}^1 ({\cal H})$ the space of trace class operators
${\cal H}= L^2(\R^n)$. Denote by  ${\cal D}$ the space of operators
$A$ in ${\cal L}^1 ({\cal H})$ such that  the  following limit
$$ \lim _{t\rightarrow 0} { e^{it\Delta} A  e^{-it\Delta} -A \over
t}  $$
exists in  ${\cal L}^1 ({\cal H})$. This  limit  is denoted by
$i[\Delta , A]$. A classical solution of  (TDHF) (for a fixed $h>0$
) is a map  $t \rightarrow \rho_h(t)$ in  $C^1(\R , {\cal L}^1
({\cal H}) \bigcap C (\R , {\cal D})$ verifying  (1.4). In [BdPF1]
and [BdPF2],  it is shown  that if  $\rho_h (t)$ is a classical
solution of (TDHF) and if  the operator $\rho_h(0) $ is  $\geq 0$,
then for all $t\in \R$, one has $\rho_h (t) \geq 0$ for all $t$, and
the trace of $\rho_h(t)$ is  constant.

\bigskip

Our first aim  is to understand  the  Weyl  symbol of
$\rho_h(t)$, whose definition is recalled in section
2. The  Weyl calculus  establishes a  bijection, that depends on a
parameter  $h>0$, between  tempered distributions on
$\R^{2n}$ (called  symbols), and the continous operators of
${\cal S}(\R^n)$ in  ${\cal S}'(\R^n)$.  For every  tempered distribution
 $F$ on  $\R^ {2n}$, and for every  $h>0$, let
$Op_h^{weyl}(F)$ be the associated Weyl operator (see  section 2).
For every continous operator $A$ of ${\cal S}(\R^n)$ in ${\cal
S}(\R^n)$, and,  for every  $h>0$, let  $\sigma_h^{weyl}(A)$ be the Weyl symbol
of  $A$. Hence,  the   equalities  $F= \sigma_h^{weyl}(A)$ and
$A = Op_h^{weyl} (F)$   are  equivalent.

\bigskip

This  work  uses  results  of C. Rondeaux [R],  that studies
the  Weyl  symbols of operators in the Schatten classes,
and in particular of trace class operators. We will denote by $W^{m ,
p} (\R^{2n})$ ($1 \leq p \leq \infty$)  the space of  functions
  that belong to  $L^p(\R^{2n})$,
  as well as all their derivatives of order $\leq m$.
It is proved in  C. Rondeaux [R] that every function in
 $W^{2n+2, 1} (\R^{2n})$ is the Weyl symbol of a  trace class operator.
One can find  in  [R] a characterization of
 trace class operators  whose Weyl  symbol is in
$W^{\infty  , 1} (\R^{2n})$.  It is also shown in  [R]  that a
trace class operator $A$ enjoys this propriety  if and only if its
iterated commutators with the following  operators
$$ P_j (h) = {h\over i} {\partial \over \partial x_j} \hskip 3cm Q_j
(h) = x_j  \leqno (1.5)$$
are  trace class operators in  ${\cal H} = L^2(\R^n)$.
 This characterization is the analogue of the one given by
Beals [B] for  bounded  operators with their Weyl symbol   in
$W^{\infty  , \infty } (\R^{2n})$. One may find in  [R] examples of
trace class operators whose Weyl 's  symbol  is not in
$L^1(\R^{2n})$, or of functions in $L^1(\R^{2n})$ that are not the
Weyl  symbol  of a trace class operator. (The first example will be
given in a more explicit form in the second part of this work.)
These results will be recalled in section 2, by making clear the
dependance on the parameter $h$ in view of semi-classical
applications, more particularly  in the proposition 2.1.

\bigskip

Our first main result is the following.

  \bigskip

{\bf Theorem  1.1.} {\it Let  $(\rho_h (t))_{(h>0)}$ be a family of
classical  solutions of (TDHF), with $V$ and  $W$ in $W^{\infty
\infty }(\R^n)$.  We assume  that, for every  $h>0$,  the operator
$\rho_h(0)$ is self adjoint  $\geq 0$, and that its trace equals 1,
that, for every  $h>0$, its Weyl  symbol  $\sigma _h^{weyl} (\rho_h
(0))$ is in  $W^{\infty 1}(\R^{2n})$, and that the family of
functions
$$ F_h = (2\pi h)^{-n}\sigma _h^{weyl} (\rho_h (0))\leqno (1.6)$$
is bounded in  $W^{\infty 1}(\R^{2n})$ independently of $h$ in  $(0,
1]$. Then, for every  $t\in \R$, the  Weyl symbol of the operator
$\rho_h(t)$ is in  $W^{\infty 1}(\R^{2n})$, and the family of
functions $u_h(., t)$ defined by:
$$ u_h(., t) = (2\pi h)^{-n}\sigma _h^{weyl} (\rho_h (t))\leqno (1.7)$$
is bounded  in  $W^{\infty 1}(\R^{2n})$ independently  of $h$
in  $(0, 1]$ and of  $t$ in a  compact set  of $\R$.

 }

\bigskip

The  hypothesis of  positivity  of  $\rho_h(0)$ serves to ensure
that  the trace norm of  $\rho_h(t)$ remains  constant.  Without
this  hypothesis it is the trace that   stays constant, and not the
trace norm. This hypothesis  is verified if  $\rho_h(0) = (2\pi h)^n
Op_h^{AW}(G)$, with $G \geq 0$ in  $L^1(\R^{2n})$, where
$Op_h^{AW}(G)$ is the anti-Wick operator  associated  to $G$ (see
section 2).  One finds in  Bove da Prato Fano [BdPF1] and [BdPF2]
the proof of the fact
  that, if  $\rho_h (0) \geq 0$,  then  $\rho_h (t) \geq 0$  for every
 $t$. Observe that the positivity hypothesis concerns
the operator and not  the  symbol.

\bigskip

It is therefore natural to give an asymptotic expansion in terms of
powers of  $h$, of the function  $u_h(., t)$ of (1.7). The first
term will be a solution of the Vlasov  equation  and the rest will
be estimated   in the $L^1$ norm. One can see in   A. Domps, P.
L'Eplattenier, P.G. Reinhard et E. Suraud [DLERS]  a physical
formulation  of this  problem.
 The successive terms  depend on the initial data (1.6).   If the  data  depends on $h$,
without possessing an asympotic expansion as powers of  $h$, it is
inevitable that the  successive terms in the  expansion de $u_h(., t)$  depend  on $h$.

\bigskip

{\bf Theorem  1.2.} {\it Let  $(\rho_h (t)$ be a classical solution
of  (TDHF) verifying the  hypotheses of theorem
1.1. Then there exists a  sequence of functions $(X , t) \rightarrow
u_j (X , t, h)$ on  $\R^{2n} \times \R$ ($j\geq 0$), depending  on the
parameter $h>0$, such that

\smallskip

- The function  $t \rightarrow u_j (., t, h)$ is  $C^{\infty}$ on
$\R$ valued in  $W^{\infty, 1}(\R^{2n})$. For every
multi-index $(\alpha , \beta)$, there exists  a function $C_{\alpha
\beta }(t)$, bounded on every compact set  of $\R$, such that:
$$\Vert \partial_x^{\alpha} \partial _{\xi}^{\beta} u_j (. , t, h)
\Vert _{L^1 (\R^{2n})} \leq C_{\alpha \beta }(t)   \leqno (1.8)$$
for all  $t \in \R$ and  $h\in (0, 1]$.

  \smallskip

-  One has, if  $F_h$ is the  function defined in  (1.6):
$$ u_0 (X , 0, h) = F_h (X)  \hskip 2cm u_j (X , 0, h) = 0. \ \ \ \ \  (j\geq 1) \leqno (1.9)$$

  \smallskip

- The function $u_0 (X , t , h)$ verifies  the Vlasov  equation:
$${\partial u_0  \over \partial t} + 2 \sum _{j=1}^n \xi_j {\partial
u_0  \over \partial x_j} = \sum _{j= 1}^n {\partial \over \partial
x_j} V_{cl} (u_0 (., t)) \  {\partial u_0 (., t)\over \partial
\xi_j}. \leqno (1.10)$$

 \smallskip

-  For every $N\geq 1$, the function $u_h (., t)$ defined  in (1.7)
and the function $F^{(N)} (., t , h) $ defined  by:
$$  F^{(N)} (X , t , h) = \sum _{k= 0}^{N-1} h^j u_j (X , t, h)\leqno (1.11) $$
verify  if $h\in (0, 1]$:
$$ \Vert u_h (., t) - F^{(N)} (., t , h)
\Vert _{L^1(\R^{2n})} \leq C_N (t) h^N, \leqno (1.12) $$
where $C_N $ is a  function on $\R$, bounded on  the  compact sets.

 \smallskip

 -For every  $N\geq 1$, the operator $\rho_h ^{(N)} (t)$ defined
 by:
$$ \rho_h^{(N)} (t) = (2\pi h)^n \ Op_h^{weyl} ( F^{(N)} (., t
, h)), \leqno (1.13) $$
(where  $  F^{(N)} (. , t , h)$ is the function of (1.11)),
verifies:
$$\Vert  \rho_h(t) -  \rho_h^{(N)}(t) \Vert _{{\cal L}^1({\cal H})}
\leq C(t) h^{N+1}.   \leqno (1.14) $$

}

\bigskip

 In section 5, we will make precise the  construction
of the $u_j $ ($j\geq 1$), and we will prove the theorem.

\bigskip

In the absence  of  mutual interaction ($W = 0$), and if one studies
the time evolution of bounded operators instead  of trace class
operators, the time evolution  of the  Weyl symbol  is described  by
the Egorov theorem (see [H], [R1], [BR]).

\bigskip

If one only makes the  hypothesis that  $\rho_h (0)$ is trace class,
the Weyl symbol  of $\rho_h (0)$ is well  defined, but is not
necessarily in $L^1(\R^{2n})$,  which is the natural class to
compare with a  solution of the Vlasov equation A counterexample is
outlined  in  [R]  and will be detailled in the second part of this
work. If one wants to strongly weaken the hypotheses on the initial
data  $\rho_h(0)$, the  Weyl calculus is not anymore the right
setting to establish a link with the Vlasov equation.
 On the other hand, we will see in  a second part of this  work that, under
 much  weaker hypotheses  than the ones
of theorems 1.1 and 1.2, the  Wick symbol of the solution can be
used for the relationship between the two equations (TDHF) and
Vlasov.

  \bigskip

In section  2, we will recall the results of  C. Rondeaux [R] by
making precise  the dependance on the  parameter $h$ in view of applications
to semi classical  analysis. The corresponding   results on the
composition of symbols and  the Moyal bracket  are stated
in  section 3. The  results of sections 2 et 3 are proved in the
 appendixes A et B.  For section 2 and
appendix  A  we  use a technique of  A. Unterberger [U1] and [U2].
The sections 4 et 5 are  dedicated  to the proofs  of theorems 1.1
and 1.2.

\bigskip

{\bf 2.   Weyl calculus and trace class operators.}

\bigskip

We  denote by   ${\cal H} = L^2(\R^n)$   and  by $ {\cal L}^1({\cal
H})$ the set of  trace class operators in ${\cal H}$. This space is
a normed space  with  the norm defined by :
$$\Vert A \Vert _{{\cal L}^1 ({\cal H})} =  Tr \Big ( (A^{\star}A)^{1/2}  \Big
).
\leqno (2.1)$$
We will denote  by $W^{m , p} (\R^{2n})$ ($1\leq p \leq +\infty$,
$m$ integer $\geq 0$ or $m= +\infty$)
 the space of functions $F$  which are in
$L^p(\R^{2n})$,  as well as all their derivatives up to order  $m$.

\bigskip

Weyl calculus   sets a  bijection between  the operators $A$ of
${\cal S}(\R^n)$  in  ${\cal S}'(\R^n)$, thus admitting a
distribution kernel in ${\cal S}'(\R^{2n})$, and tempered
distributions   on  $\R^{2n}$ (symbols). This bijection  may depend
on  a parameter  $h>0$.  For every  $F $ in  ${\cal  S}'(\R^{2n})$,
we set
 $Op_h^{weyl} (F)$  the operator  $A: {\cal S}(\R^n) \rightarrow {\cal S}'(\R^n)$
 whose distribution kernel
is:
$$ K_A (x , y) = ( 2 \pi h)^{-n} \int _{\R^{n}} F({x+y \over 2} , \xi )e^{{i\over h}
(x-y).\xi} d\xi.  \leqno (2.2)$$
This  relationship, understood in the sense of distributions, may be
inverted.  We will denote $\sigma_h^{weyl} (A)$  the distribution
$F$ (symbol of  Weyl of $A$)  such that $A = Op_h^{weyl}(F)$.
 In view of applications to trace class operators we can rewrite (2.2) equivalently
when  $F$ is in  $L^1(\R^{2n})$  as:
$$ Op_h^{weyl} (F) = (\pi h)^{-n}  \int_{\R^{2n}} F(X) \Sigma _{Xh}
dX, \leqno (2.3)$$
where, for  $X = (x , \xi)$ in $\R^{2n}$,   $\Sigma _{X h}$ is  the
"symmetry"  operator  defined by:
$$ (\Sigma _{X h} f) (u) = e^{{2i\over h} (u-x)\xi } f( 2x - u)
\hskip 1cm  X = (x , \xi)\in \R^{2n}. \leqno (2.4) $$
If  $A$ is  trace class, one has
$$ \sigma_h^{weyl} (A) (X) = 2^n Tr ( A \circ \Sigma_{Xh}) \hskip 2cm X \in
\R^{2n}. \leqno (2.5) $$

According to  Calderon-Vaillancourt [CV] (see also H\"ormander [H])
if $F$ is in  $W^{\infty \infty }(\R^{2n})$, then $Op_h^{weyl}(F)$
is bounded in $L^2(\R^n)$. If $F$ is in $L^1(\R^{2n} )$ the operator
$Op_h^{weyl}(F)$  is bounded by (2.3)  but not necessarily trace
class. It is shown in   C.Rondeaux [R] that, if  $F$ is in  $W^{m,
p} (\R^{2n})$, ($1\leq p < \infty$, $m$  large enough),  the
operator $Op_h^{weyl}(F)$ is in the Schatten class ${\cal L}^p({\cal
H})$. For $p=1$, we  will precisely state this in  proposition 2.1.
If $F$ is in  $W^{\infty, 1} (\R^{2n})$, one has:
$$ Tr ( Op_h^{weyl} (F) = (2 \pi h)^{-n} \int _{\R^{2n}} F(X) dX.  \leqno (2.6) $$
If  $F$ is  in $W^{\infty p}(\R^{2n})$ and  $G$ in $W^{\infty
q}(\R^{2n})$ ($p\geq 1$, $q\geq 1$, ${1\over p} + {1 \over q}=1$),
one has
$$ Tr \Big ( Op_h^{weyl}(F) \circ Op_h^{weyl}(G) \Big ) =
(2 \pi h)^{-n} \int _{\R^{2n}} F(X)  G(X) dX.  \leqno (2.7) $$
The left hand side  makes sense,  since from  [R],  the two
operators  under composition  are respectively ${\cal L}^p({\cal
H})$ and  ${\cal L}^q({\cal H})$, in such a way that  their
composition is trace class.

\bigskip

The next proposition  mainly due  to  C. Rondeaux [R], gives
sufficient conditions   (on the  Weyl   symbol) for an operator to
be  trace class,
 and gives  a characterization of the set of operators whose Weyl symbol is in
$W^{\infty 1} (\R^{2n})$. This characterization is the analogue of
the  Beals characterization  [B], which  concerns the symbols in  $W^{\infty
\infty} (\R^{2n})$.

 We note $P_j (h) = {h\over i} {\partial \over
\partial x_j}$
the momentum operators and we note $Q_j (h)$   the multiplication by $x_j$.

For every operator  $A$  of  ${\cal  S}(\R^n)$  in  ${\cal S}'(\R^n)$,  we set
$$(ad\  P(h) )^{\alpha} (ad \ Q(h))^{\beta } A =(ad \
P_1 (h))^{\alpha_1} ...(ad \ Q_n (h))^{\beta _n}A. \leqno (2.8) $$

\bigskip

{\bf Proposition 2.1.} {\it a) If  $F$ is in $L^1(\R^{2n})$ as well  as all
its derivatives up to order
 $2n+2$, then, for all  $h>0$, the operator  $Op_h^{weyl}(F)$ is trace class, and:
$$ \Vert Op_h^{weyl}(F) \Vert _ {{\cal L}^1({\cal H})} \leq C
h^{-n} \sum _{|\alpha |+|\beta|\leq 2n+2} h^{(|\alpha |+|\beta |)/2}
\Vert \partial _x^{\alpha} \partial _{\xi}^{\beta}F \Vert
_{L^1(\R^{2n})}.  \leqno (2.9)$$
b) If  $A$ is a trace class operator, and if, for every multi-index
$(\alpha , \beta)$ such that  $|\alpha|+ |\beta|\leq 2n+2$, the
operator  $(ad\  P(h) )^{\alpha} (ad \ Q(h))^{\beta } A $ is trace
class, then the Weyl  symbol of  $A$  is in $L^1(\R^{2n})$   and:
$$ (2\pi h)^{-n}  \Vert \sigma_h^{weyl} (A)\Vert _{L^1(\R^{2n})}
\leq C \sum _{|\alpha |+ |\beta| \leq 2n+2} h^{-(|\alpha|+
|\beta|)/2} \Vert (ad\  P(h) )^{\alpha} (ad \ Q(h))^{\beta } A \Vert
_{{\cal L}^1 ({\cal H})}, \leqno (2.10)$$
where the constant  $C$ depends only on  $n$.

\smallskip

c) the following are equivalent :
\smallskip
  i) A family of  operators $(A_h)_{0< h \leq 1}$ is  of the form
  $A_h = Op_h^{weyl}(F_h)$,  where
  $(F_h)$ is a bounded family of  functions in  $W^{\infty 1}(\R^{2n})$.
  \smallskip
  ii) For every  $h>0$,  the operator $A_h$  is trace class , as well as all iterated
  commutators of  $A_h$   with the operators  $P_j(h)$    and
  $Q_j(h)$    and ,  for every $(\alpha , \beta)$,  the family of  norms
  $$ h ^{n - |\alpha|-|\beta|}  \ \Vert (ad\  P(h) )^{\alpha} (ad \ Q(h))^{\beta } A \Vert
_{{\cal L}^1 ({\cal H})}  \leqno (2.11)$$
  stays bounded  when  $h$ varies  in  $(0, 1]$.

 }

\bigskip

Parts  a)  and    b)  are proved  in C. Rondeaux [R],  without the
parameter $h$. Part a)  needs only minor modifications  to introduce
this parameter but it seems to us better to give a proof of Part b)
in appendix A.

For the sake of clarity  it might be useful to recall  the
well-known analogue of proposition 2.1 for  the bounded operators
and the symbols  in   $W^{\infty \infty}(\R^{2n})$, that is to say,
the  Calderon-Vaillancourt theorem and the Beals  characterization.

\bigskip

{\bf Proposition 2.2.} {\it a) If  $F$ is in $L^{\infty} (\R^{2n})$
as well as all derivatives up to order $2n+2$, then, for all  $h>0$,
the operator  $Op_h^{weyl}(F)$  is bounded in ${\cal H} =
L^2(\R^n)$,   and:
$$ \Vert Op_h^{weyl}(F) \Vert _ {{\cal L}({\cal H})} \leq C
 \sum _{|\alpha |+|\beta|\leq 2n+2} h^{(|\alpha |+|\beta |)/2}
\Vert \partial _x^{\alpha} \partial _{\xi}^{\beta}F \Vert
_{L^{\infty} (\R^{2n})}. \leqno (2.12) $$
\smallskip

b) If  $A$ is a bounded operator and if, for all multi-indices
$(\alpha , \beta)$ such that  $|\alpha|+ |\beta|\leq 2n+2$, the
operator $(ad\ P(h) )^{\alpha} (ad \ Q(h))^{\beta } A$   is bounded,
then the Weyl symbol  of $A$  is in  $L^{\infty }(\R^{2n})$, and one
has:
$$ \Vert \sigma_h^{weyl} (A)\Vert _{L^{\infty} (\R^{2n})}
\leq C \sum _{|\alpha |+ |\beta| \leq 2n+2} h^{-(|\alpha|+
|\beta|)/2} \Vert (ad\  P(h) )^{\alpha} (ad \ Q(h))^{\beta } A \Vert
_{{\cal L}({\cal H})}. \leqno (2.13)$$

}

\bigskip

{\it  Anti-Wick calculus.}  The  definition of this calculus  uses  coherent states.
By this  we mean  the  family  of functions $\Psi _{Xh }$ in $L^2 (\R^n)$, indexed  by  the  parameter  $X = (x , \xi)
\in \R^{2n}$, and  depending on  $h>0$,  defined by
$$  \Psi_{X, h } (u) =  (\pi h)^{ -n/4}e^{-{| u-x|^2 \over
2h}} e^{{i \over h} u .\xi - {i \over 2h} x. \xi} \hskip 1cm X = (x
, \xi)\in \R^{2n}. \leqno (2.14)$$
These functions, called coherent states,  will be used  to give
examples of operators verifying  the hypotheses of theorem  1.1, and
will be also helpful   to prove  Proposition  2.1 (appendix A). Fore
more properties of the coherent states and the anti-Wick calculus,
see, for instance [CF], [F], [CR1], [L1],  [R1 ], [U1], [U2].  The
two fundamental properties are the fact  that:
$$ | < \Psi_{Xh} , \Psi_{Yh}>| = e^{- {1 \over 4h} |X-Y|^2} \hskip  3cm \Vert \Psi_{Xh}\Vert =
1,
\leqno (2.15)$$
and that, for all  $f$ and  $g$   in   ${\cal H}$
$$ < f , g> = (2 \pi h)^{-n} \int _{\R^{2n}} < f , \Psi _{X h} > \ <
\Psi _{X  h} , g > \ dX .   \leqno (2.16)$$
 For every function  $F$ in $L^1(\R^{2n})$  and for every
$h>0$  we set  $Op_h^{AW} (F)$   to be   the bounded operator  in $L^2(R^n)$
such  that  for all   $f$  and $g$  in  ${\cal H}$:
$$ < Op _h ^{AW}(F) f , g >  =  ( 2 \pi h)^{-n} \int_{\R^{2n}} a(X)
< f , \Psi_{Xh}> \ < \Psi_{Xh}, g>
 \ dX.   \leqno (2.17) $$
If   $F$ is in  $L^1(\R^{2n})$ we see  that  $Op _h ^{AW}(F)$  is
indeed trace class  in  ${\cal H}$,  and that:
$$  \Vert Op_h^{AW} (F)  \Vert _{ {\cal L}^1 ({\cal H})}  \leq  (2 \pi
h)^{-n} \int _{\R^{2n}} |F(X)| dX. \leqno (2.18 )$$
Moreover one has:
$$ Tr (Op _h ^{AW}(F)) =  (2 \pi h)^{-n} \int_{\R^{2n}} F(X) dX.
\leqno (2.19)$$
If  $F\geq 0$,  the operator  $Op_h^{AW}(F)$   is self adjoint   and   $\geq 0$.
 The  Weyl  symbol   of the operator $Op_h^{AW} (F)$   is given by :
$$\sigma _h ^{weyl} \Big ( Op_h^{AW} (F) \Big ) = e^{{h\over
4}\Delta } F,  \leqno (2.20) $$
where  $\Delta $ is  the  Laplacian  on  $\R^{2n}$.   In fact,  the
operators  $\Sigma _{Yh}$ defined in  (2.4)  and the operator
$P_{Xh}$ of   orthogonal projection  on  ${\rm Vect} (\Psi_{Xh})$
verify:
$$ Tr ( P_{Xh} \Sigma_{Yh} ) = e^{-{|X-Y|^2 \over h}}.  \leqno (2.21)$$

\bigskip

{\bf 3.  Basic  Facts on the  Moyal  bracket.  }

\bigskip

The   composition of  symbols  in the Weyl  calculus   is a very classical field
 (see  H\"ormander [H ] or  D. Robert [R1]  for the
dependance on the semiclassical parameter $h$). We need to adapt this
to the  classes of  C. Rondeaux symbols,  by making precise the dependance on the parameter $h$.

\bigskip

We define a differential operator  $ \sigma (\nabla_1 , \nabla _2)$
on $\R^{2n}\times \R^{2n}$  by:
$$  \sigma (\nabla_1 , \nabla _2) = \sum _{j=1}^n {\partial^2 \over \partial y_j \partial \xi_j} -
{\partial^2 \over \partial x_j \partial \eta_j},  \leqno (3.1)$$
where   $(x , \xi , y , \eta)$   denotes  the variable of  $
\R^{2n}\times \R^{2n}$.

\bigskip

{\bf Theorem 3.1.} {\it For all  functions $F$ in  $ W^{\infty
p}(\R^{2n})$ and $G$ in $ W^{\infty q }(\R^{2n})$, ($p\geq 1$,
$q\geq 1$, ${1\over p} + {1 \over q} = 1$), for all $h>0$,  there
exists a function $M_h (F , G, .)$  in  $W^{\infty 1 }(\R^{2n})$
(Moyal bracket) such that:
$$\Big [ Op_h^{weyl}(F)  ,  Op_h^{weyl}(G)\Big ]  =
Op_h ^{weyl}(M_h (F , G, .) ). \leqno (3.2) $$
For all  integers  $N \geq 2$, one has:
$$ M_h(F , G, X) = \sum _{k=1}^{N-1} h^k C_k (F , G , X)
 + R^{(N)}_h (F , G , X ), \leqno (3.3)$$
where  the  function $C_k(F , G , X)$   is  defined   by:
$$C_k(F , G , X) = {1 \over (2i)^k k!}  \Big [   \sigma
(\nabla_1 , \nabla _2)^k ( F \otimes G)  (X , X)  -  \sigma
(\nabla_1 , \nabla _2)^k ( G \otimes F )  (X , X)\Big ], \leqno
(3.4)
$$
and where  the  function $R^{(N)}_h (F , G , .)$  is in  $W^{\infty
1}(\R^{2n})$.  We can write also (3.3) for $N= 1$, with the
convention that the sum vanishes, and that $R^{(N)}_h  = M_h$.  For
every integer $\ell$, there exists a constant $C$ such that:
$$h^{\ell /2} \Vert \nabla ^{\ell} R ^{(N)}_h (F , G , .) \Vert _{L^1(\R^{2n})}  \leq C
 \sum _{
 |\alpha | + |\beta |\leq \ell + 4n+2 + 2N  \atop |\alpha | \geq N, |\beta| \geq N }
 h^{(\alpha + \beta )/2}   \Vert  \nabla ^{\alpha} F
\Vert _{L^p (\R^{2n})} \ \Vert  \nabla ^{\beta} G \Vert _{L^{q
}(\R^{2n})}.  \leqno (3.5) $$
The operator  $\widehat R ^{(N)}_h (F , G )= Op_h^{weyl} (R ^{(N)}_h
(F , G , .))$ verifies:
$$ \Vert \widehat R ^{(N)}_h (F , G) \Vert _{{\cal L}^1({\cal H})} \leq C
h^{-n}  \sum _{
 |\alpha | + |\beta |\leq 6n+4 + 2N  \atop |\alpha | \geq N, |\beta| \geq N }
 h^{(\alpha + \beta )/2}   \Vert  \nabla ^{\alpha} F
\Vert _{L^p (\R^{2n})} \ \Vert  \nabla ^{\beta} G \Vert _{L^{q
}(\R^{2n})}.  \leqno (3.6) $$

 }

\bigskip

This theorem will be  proved  in  the appendix  B.  We will also use in the second part of this
work the well known analogue  of  theorem 3.1, which we recall here in order to allow
for its application when  needed.

\bigskip

{\bf Theorem 3.2.} {\it With the notations of  theorem 3.1,
if   the  functions $F$ et $G$  are in  $ W^{\infty \infty}(\R^{2n})$,
the function $R_h^{(N)} (F , G , .)$ defined  by the  equality ()
verify,  for any $\ell \geq 0$:
$$ h^{\ell /2} \Vert \nabla ^{\ell} R ^{(N)}_h (F , G , .) \Vert _{L^{\infty}(\R^{2n})}
 \leq C \sum _{ j\geq N, k\geq N \atop \ell + 2N \leq j + k \leq \ell + 2N + 4n+2}
 h^{(j+k)/2} \Vert \nabla ^j F \Vert _{L^{\infty}(\R^{2n})} \
 \Vert \nabla ^k G \Vert _{L^{\infty}(\R^{2n})}\leqno (3.7)$$
The operator $ \widehat R ^{(N)}_h (F , G)$ verify:
$$ \Vert \widehat R ^{(N)}_h (F , G) \Vert _{{\cal L}({\cal H})}
\leq C \sum _{ j\geq N, k\geq N \atop  2N \leq j + k \leq  2N +
6n+4} h^{(j+k)/2} \Vert \nabla ^j F \Vert _{L^{\infty}(\R^{2n})} \
 \Vert \nabla ^k G \Vert _{L^{\infty}(\R^{2n})} \leqno (3.8) $$

}
\bigskip

{\bf 4. Egorov's theorem  for trace class operators  and proof  of
theorem 1.1. }

\bigskip

We are going to adapt to the case of symbols in $L^1(\R^{2n})$  and
trace class operators the idea of the proof of  Egorov's theorem
contained in the book of  D. Robert [R1].  The difference with [R1]
comes from the fact  the classes of operators considered here are
the classes introduced by C.  Rondeaux and that  Hamiltonians are
time dependent.

\bigskip

We consider a function $(x , t ) \rightarrow V(x , t)$ on $\R^n
\times \R$, real valued, depending in a $C^{\infty } $ manner  on
$x$,  and continuously  on $t$.  We suppose that, for every
$\alpha$, there exists $C_{\alpha}>0$  such that:
$$ |\partial_x^{\alpha} V(x , t) |\leq C_{\alpha} \hskip 2cm (x , t)
\in \R^n \times \R.  \leqno (4.1)$$
We set:
$$ H (x , \xi , t) = |\xi|^2 + V  (x , t).  \leqno (4.2)$$
We  denote by $V(t)$    the multiplication by  $V(., t)$.  We set:
$$ \widehat H_h (t) = - h^2 \Delta + V(t).  \leqno (4.3)$$
Therefore, $\widehat H_h (t) = Op_h^{weyl} (H(., t))$.  Let us
recall now some facts on unitary propagators ([RS]).

\bigskip

{\bf Proposition 4.1.} {\it For all $t\in \R$,  let  $V(., t)$  a
$C^{\infty }$ function  on  $\R^n$, verifying  (4.1),  and depending
in a $C^1$ manner on $t$. Let $ \widehat H_h (t)$ be the operator
defined in  (4.3). For every $f$ in ${\cal S}(\R^n)$, and for every
$s\in \R$, there exists a function denoted by $t \rightarrow U_h(t ,
s)f$ that verifies:
$$ i h {\partial \over \partial t} U_h(t , s)f = ( \widehat H_h (t))U_h(t ,
s)f, \hskip 2cm U_h(s , s)f = f.  \leqno (4.4)$$
The operator  $U_h(t , s)$ maps  ${\cal S}(\R^n)$ into itself and,
by duality, ${\cal S}'(\R^n)$   into itself.  One  has $U_h (s , t)
= U_h(t , s)^{-1}$. One also  has:
$$i h {\partial \over \partial s} U_h(t , s) = - U_h(t , s)
( \widehat H_h (s)).  \leqno (4.5) $$
For every  operator  $A$  of  ${\cal S}(\R^n)$  into  ${\cal
S}'(\R^n)$,  let us set:
$$ G_h(t , s) (A) = U_h(t , s) \circ A \circ U_h(s , t). \leqno (4.6)$$
One has
$$i h {\partial \over \partial t} G_h(t , s) (A) =
\Big [ \widehat H_h (t) , G_h(t , s) (A) \Big ] \hskip 2cm G_h(s ,
s) (A) = A.   \leqno (4.7)$$
}

\bigskip

The analogue of Egorov's theorem for operators in the class of [R]
is the following.

\bigskip

{\bf Theorem 4.2.} {\it Let  $F$  be a function  defined on
$W^{\infty 1}(\R^{2n})$. Let  $A_h = Op_h^{weyl}(A)$. Then, for
every  $t\in \R$, there exists  a function  $F_{h t}$  in $W^{\infty
1}(\R^{2n})$ such that:
$$ G_h (t , 0) ( A_h) = Op_h^{weyl} (F_{h t}).   \leqno (4.8)$$
If the function $F$ and the potential $V(., t)$ depends   on a
parameter  $\lambda$, while staying  bounded respectively  in
$W^{\infty 1}(\R^{2n})$  and  in $W^{\infty \infty}(\R^{2n})$
independently  of  $\lambda$, (and of $t$ for $V$),  then  the
function $F_{h t}$ remains bounded  in  $W^{\infty 1}(\R^{2n})$
independently of $\lambda$,  of $h$ in  $(0, 1]$, and of  $t$ in a
 compact set of  $\R$.

}

\bigskip

Following the idea  of  [R1], which is related in some sense  to
Dyson's series,  we will express  our solution  $G_h (t , 0) ( A_h)$
under the form:
$$ G_h (t , 0) ( A_h)  = \sum _{k= 0}^{N-1} Op_h^{weyl } \Big ( D_k (., t)
\Big ) + h^N E_N (t , h), \leqno (4.9) $$
where the  functions $D_k (., t)$   are  in $W^{\infty 1}(\R^{2n})$  and
$E_N (t , h)$ will  be a  trace class operator  with bounded trace norm.
 In  a second step,  we will  show that the  commutators of  $E_N(t , h)$  with the position and
momentum operators  are  trace class operators, and we will estimate
their traces. Finally  we will rely on the characterization of C.
Rondeaux (recalled in  proposition 2.1)  to show that  $ G_h (t , 0)
( A_h)$  is itself a  pseudodifferential  operator, with  a symbol
in $W^{\infty 1}(\R^{2n})$.

\bigskip

The construction of  the  terms  $D_k(., t)$ will use the
Hamiltonian flow  of  $H(., t)$.  For every function  $G$  in  $ W^{\infty
1}(\R^{2n})$, we call  $\Phi_{t s } (G)$  the function on $\R^{2n}$
defined  by
$$ {\partial \Phi_{t , s}  (G) \over \partial t} = \{  H(., t) , \Phi_{t s}(G) \}, \hskip 2cm
\Phi_{s , s} (G) = G.  \leqno (4.10) $$
Under hypothese  (4.1), one knows that  if  $G= G_{\lambda}$  where
$(G_{\lambda})_{\lambda \in E}$ is a family  of   bounded functions
in $ W^{\infty 1 }(\R^{2n})$,  then  $\Phi_{t s}
(G_{\lambda})$  stays bounded  $W^{\infty 1 }(\R^{2n})$   when
$(t , s)$ varies in a  compact set  of  $\R^2$  and  $\lambda$ in  $E$.

\bigskip

{\bf Lemme 4.3.} {\it  For every  function   $G$  in  $ W^{\infty 1
}(\R^{2n})$  and for  $(t , s) $ in  $\R^2$, one has
$$G_h (t , s) \Big (OP_h^{weyl} (G)\Big ) =  Op_h^{weyl} ( \Phi_{t s} (G) )
+
 h  \int _s^t G_h (t, t_1)
\Big ( Op_h^{weyl} ( R(., t_1 , s, h)) \Big ) \  dt_1,  \leqno
(4.11)$$
where the  function  $R(., t_1 , s, h)$  is in  $ W^{\infty 1
}(\R^{2n})$. If  $G$  belongs   to  a family  of  bounded  functions
in  $ W^{\infty 1 }(\R^{2n})$, it is also the case for  $R(., t_1 ,
s, h)$ when  $(t_1 , s)$ lies in a compact set  of $\R^2$ and $h$
is in  $(0, 1]$.

}

\bigskip

{\it Proof of the  lemma.}   From definition (4.10):
$$ {\partial  \over \partial t} Op_h^{weyl} (\Phi_{t s} (G) ) =
Op_h^{weyl} (  \{  H(., t) , \Phi_{t s }(G) \}).  $$
With the notations  of theorem 3.1, with  $N=2$, one  may write:
$$ [ \widehat H_h( t) , Op_h^{weyl} (\Phi_{t s h}(G)  )] =
{h\over i}Op_h^{weyl} (  \{  H(., t) , \Phi_{t s }(G) \}) +
OP_h^{weyl}\Big ( R_h^{(2)}( H(., t)\  , \Phi_{t s}(G))\Big ).
$$
Consequently:
$$ {\partial  \over \partial t} Op_h^{weyl} (\Phi_{t s } (G) )
- {i \over h} [ \widehat H_h( t) , Op_h^{weyl} (\Phi_{t s }(G)  )]
=- {i \over h}OP_h^{weyl}\Big ( R_h^{(2)}(H(., t) , \Phi_{t s }(G))
\Big ). $$
On the other hand,
$$ {\partial  \over \partial t}G_h (t , s) (OP_h^{weyl} (G))
- {i \over h}  [ \widehat H_h(t)  , G_h (t , s) (OP_h^{weyl} (G)) ]
= 0$$
By combining  these two equalities,  noting that  for $t=s$, the two
operators $ G_h (s , s)  ( Op_h^{weyl}(G))$    and $Op_h^{weyl}(
\Phi _{s s }(G))$  are equal,  and then  by using Duhamel ' s
principle,  we obtain  (4.11),  with:
$$R(., t_1 , s, h) = - {i\over h}R_h^{(2)}\Big ( H (., t_1) , \Phi_{t_1 s }(G)
\Big ).  $$
It is well  known that, when  $F(x , \xi) = |\xi|^2$,   one has
$R_h^{(2)}\Big ( F , G \Big )  = 0$  for every function $G$.  Hence:
$$R(., t_1 , s, h) = - {i\over h^2}R_h^{(2)}\Big ( V (., t_1) , \Phi_{t_1 s }(G)
\Big ). $$
By hypothesis,  $V(., t_1)$ is in  $W^{\infty \infty }(\R^n)$ and is
bounded, independently  of $t_1$.   We have seen that  $\Phi_{t_1 s
}(G)$ is in  $W^{\infty 1 }(\R^{2n})$, independently  bounded of
$t_1$ and of $s$   when  $(t_1 , s)$ varies in a compact set of
$\R^2$. According to  theorem 3.1,  applied to the case   $N= 2$, it
follows that  $R(., t_1 , s, h)$  is  in  $W^{\infty 1 }(\R^{2n})$,
independently bounded of  $t_1$ and of  $s$    when  $(t_1 , s)$
varies  in a  compact set  of  $\R^2$  and  $h$   in  $(0, 1]$.

\bigskip

{\it Proof  of theorem  4.2.  First step. }  Let $F$ be a function
in $W^{\infty 1}(\R^{2n})$. Let  $A_h = Op_h^{weyl}(F)$. Let
$\Phi_{t s } (G)$ be  the function verifying  (4.10). For every
$t\in \R$, we define a function $D_0 (t) (.)$  in  $ W^{\infty 1
}(\R^{2n})$  by
$$ D_0(., t) =    \Phi_{t , 0}(F).  \leqno (4.12) $$
We have seen that this  function is  in  $W^{\infty 1}(\R^{2n})$,
 independently  bounded  of  $t$   on  every  compact set  of $\R$.
By   lemma  4.3 applied  to $G= F$ and  $s=0$,  and from (4.12), one
has:
$$ G_h (t , 0) ( A_h) =   Op_h^{weyl} ( D_0(t) )  + h  \int _0^t G_h (t, t_1)
\Big ( Op_h^{weyl} ( R_1(., t_1 ,  h)) \Big ) \  dt_1, $$
where $R_1(., t_1 ,  h)$   stays bounded in $W^{\infty 1 (\R^{2n})}$
when  $t_1$  belongs to a compact set   of $\R$  and  $h$  is in
$(0, 1]$. We   reiterate, by applying  4.3  with  $s= t_1$ and $G=
R_1(., t_1 ,  h)$.  We obtain:
$$ G_h (t, t_1) \Big ( Op_h^{weyl} ( R_1(., t_1 ,  h)) =
Op_h^{weyl} \Big ( \Phi (t , t_1) ( R_1 (., t_1 , h)) \Big ) + h
\int _{t_1}^t G_h ( t , t_2) (Op_h^{weyl} ( R_2 (., t_2 , t_1 , h)))
dt_2,$$
where   $R_2(., t_2 , t_1 ,  h)$ stays bounded in  $W^{\infty 1
}(\R^{2n})$ when $(t_1 , t_2) $ belongs to  a compact set of  $\R^2$
and $h$  is in $(0, 1]$. We   define a  function $D_1(., t)$ in
$W^{\infty 1 }(\R^{2n})$ by:
$$D_1(., t) = \int _0^t \Phi (t , t_1) ( R_1 (., t_1 , h)) dt_1. $$
This function is in  $W^{\infty 1}(\R^{2n})$, independently bounded
of $t$ on very compact set  of  $\R$. One has, if $t>0$:
$$ G_h (t , 0) ( A_h) = Op_h ^{weyl} \Big (D_0(t) + hD_1(t) \Big ) + h^2 \int _{ 0< t_1
< t_2 < t} G_h ( t , t_2) \Big ( Op_h^{weyl} ( R_2 (., t_2 , t_1 ,
h)) \Big )  dt_1 \ dt_2. $$
By reiterating  this process, we  obtain, for every  $N$, the
equality (4.9), with:
$$ E_N(t ,  h) = \int _{\Delta_N(t , 0)} G_h ( t , t_N) \Big (   Op_h^{weyl} (R_N
(., t_N , ... , t_1 , h)) \Big ) \ dt_1 ... dt_N,  \leqno (4.13)$$
where   $ \Delta _N (t , s)$ is the set defined,  if  $s <t$, by:
$$ \Delta _N (t , s) = \{ (t_1 , ... t_N)\in \R^N, \hskip 1cm s <
t_1 <... < t_N < t \}, \leqno (4.14)$$
and in a symmetric  way if  $s>t$. In  (4.9),  the  $D_j(., t ,
h)$ ($j\geq 0$), and  $R_N(., t_N, ..., t_1, h)$,  are in  $W^{\infty
1 }(\R^{2n})$, independently  bounded of  $h$ in $(0, 1]$, of
$(t_1 , ... t_N)$ in $\Delta_N(t , 0)$,  and of  $t$ in a compact set
of $\R$.

\bigskip
The second step will consist  in getting upper bounds for trace norms  of
iterated commutators  of $E_N(t , h)$ with the  position and momentum operators,
 with the help of a less precise technique  than the one used in the first step
  for powers  of  $h$.   However  this lesser precision will be compensated
 by the presence of the factor $h^N$ in  (4.9).  In order to do  that, we will
 use the following  lemma,
 that will be useful also in section 5 and in a second article.
 (part  II of this work). If an   operator $A$ is
 bounded  in  $L^2(\R^n)$, as well as all its iterated  commutators up
to order $m$, we set:
$$I_h^{m \infty} (A) = \sum _{|\alpha|+|\beta|\leq m} \Vert (ad \
Q(h) )^{\beta }(ad \ P(h) )^{\alpha }A \Vert _{ {\cal L}({\cal H})}.
\leqno (4.15)$$
If an   operator $A$ in  $L^2(\R^n)$ is trace class, as well as all
its  iterated commutators up to  order  $m$, we set:
$$I_h^{m, tr } (A) = \sum _{|\alpha|+|\beta|\leq m} \Vert (ad \
Q(h) )^{\beta }(ad \ P(h) )^{\alpha }A \Vert _{ {\cal L}^1({\cal
H})}.  \leqno (4.16)$$
The aim of the next lemma is to show that these properties are stable par the mapping
 $G_h(t , s)$.

\bigskip

{\bf Lemma 4.4.} {\it    Let  $\widehat H_h(t)$ be the operator
defined  in  (4.3), where  $V(., t)$ verifies  (4.1).  Let $U_h(t ,
s)$ denote  the unitary propagator and  $G_h(t , s)$ the map  of
proposition 4.1.  Let  $A$ be   a trace class operator  in ${\cal
H}=L^2(\R^n)$, as well  as all iterated  commutators $ (ad \ Q(h)
)^{\beta }(ad \ P(h))^{\alpha } A$ for $|\alpha|+|\beta|\leq m$.
Then, for all $s$ and $t$ in  $\R$, the operator  $G_h(t , s) (A)$
is also trace class, as well as all iterated  commutators with  the
$P_j(h)$ and  $Q_j(h)$  up to order  $m$. Moreover, for every
compact set  $K$ of $\R$, there exists  $C_K>0$ such that:
$$I_h^{m, tr } (G_h(t , s) (A) )   \leq C_K I_h^{m, tr } (A) \hskip
2cm (s , t)\in K^2 ,  \ \ \ \ \ h\in (0, 1].  \leqno (4.17)$$
An identical result  holds  for  bounded  operators   and for the norms  $I_h^{m, \infty }$.

}

\bigskip

We see why this lemma  alone  does not allow  to prove theorem 4.2.
One will need, for that, that in the expression (4.16),  the  terms
corresponding to  the multi-index $(\alpha , \beta)$ be multiplied
by  $h^{-|\alpha|-|\beta|}$.

\bigskip

{\it Proof of Lemma.} By  (4.7)  one checks that for every operator
 $A$ verifying  the hypothesis of the  lemma, and  for  each of the
momentum operators $P_j(h)$, the following equality:
$$ {\partial \over \partial t}  [ P_j(h) ,  G_h(t , s) (A)   ]
 - {1 \over i h} \big [ \widehat H_h(t) , [ P_j(h) ,  G_h(t , s) (A)   ] \big ] =
  {1 \over i h} \big [ [P_j(h) , \widehat H_h(t) ] , G_h(t , s) (A) \big ]. $$
Then it results, by the Duhamel  principle,  the following equality:
$$[ P_j(h) ,  G_h(t , s) (A)   ] = G_h (t , s) \Big ( [ P_j(h)  , A]
\Big ) + {1 \over i h} \int _s ^t G_h ( t , t_1) \Big ( \big [ [ P_j
(h) , \widehat H_h(t_1) ] , G_h (t_1 , s) (A) \big ] \Big ) dt_1. $$
We have  an analogous equality  for the  position operators
$Q_j(h)$. One has:
$$[P_j(h) , \widehat H_h(t) ]  =    {h\over i} {\partial V (. , t) \over \partial x_j}
 \hskip 2cm   [Q_j(h) , \widehat  H_h(t) ] = 2 i h P_j(h). $$
We therefore deduce:
$$[ P_j(h) ,  G_h(t , s) (A)   ] = G_h (t , s) \Big ( [ P_j(h)  , A]
\Big ) -  \int _s ^t G_h ( t , t_1) \Big ( \Big [ {\partial V (.,
t_1) \over \partial x_j} , G_h(t_1 , s) (A) \Big ] \Big ) dt_1,$$
$$[ Q_j(h) ,  G_h(t , s) (A)   ] = G_h (t , s) \Big ( [ Q_j(h)  , A]
\Big ) + 2  \int _s ^t G_h ( t , t_1) \Big (  \Big [ P_j(h) ,
G_h(t_1 , s) (A) \Big ] \Big ) dt_1.$$
If  $A$ and its commutators  with $P_j(h)$  and  $Q_j(h)$ are trace
class, we first observe that $[ P_j(h) ,  G_h(t , s) (A)]$ is a
trace class operator  since  $G_h (t , s)$ conserves ${\cal L}^1
({\cal H})$.  Reporting in the  second equality, we see that  $[
Q_j(h) , G_h(t , s) (A)   ]$ is itself  a  trace class operator,
 and that  the upper bound  (4.17) is proved for $m=1$.
We pursue the same reasoning to  prove (4.17), by
induction, for every  $m$. The analogue (4.17) for the
bounded operators  is proved similarly.

\bigskip

{\it Proof  of  theorem 4.2.  Second step .} Following proposition
2.1, it suffices to show  that, for every multi-index  $(\alpha ,
\beta)$ and  for every compact  set  $K$ of  $\R$,  there exists
$C_{\alpha \beta K}>0$ such that:
$$ h^{n-(|\alpha|+|\beta|)} \Vert (ad\  P(h) )^{\alpha} (ad \ Q(h))^{\beta
} G_h (t , 0) ( A_h) \Vert _{{\cal L}^1({\cal H})} \leq C_{\alpha
\beta K}, \leqno (4.18) $$
if $t\in K$  and  $h\in (0, 1]$. In order to achieve  this,  one
will use the asymptotic  expansion
 (4.9)  up to an order $N$ that will depend on  $\alpha$ and $\beta$. Since  from the first step
 the $D_j(., t , h)$ ($j\geq 0$) of  (4.9)
belong  to  $W^{\infty 1 }(\R^{2n})$, and are bounded independently
of $h\in (0, 1]$ and of  $t$ in a  compact set  of $\R$, the
proposition 2.1 shows  that:
$$ h^{n-(|\alpha|+|\beta|)} \Vert (ad\  P(h) )^{\alpha} (ad \ Q(h))^{\beta
} Op_h^{weyl} (D_j ( ., t , h))
 \Vert _{{\cal L}^1({\cal H})} \leq C_{\alpha \beta K} $$
if  $t\in K$ et $h\in (0, 1]$.  Let us derive now an upper bound
that is the analogue for the  term $E_N(t , s, h)$. For this, we
use the form  (4.13)  of  $E_N(t , s, h)$, and  we apply  lemma  4.4
with $(t , s)$ remplaced  by  $(t , t_N)$  and  $A$ by  $Op_h^{weyl}
(R_N (., t_N , ... , t_1 , h))$. Since  $R_N(., t_N, ..., t_1, h)$
is in $W^{\infty 1 }(\R^{2n})$  and is independently bounded for $h$
in $(0, 1]$, of $(t_1, ... t_N)$ in  $\Delta_N (0, t)$, and for $t$
in a compact set  of $\R$, proposition 2.1 shows that  for every
integer $m\geq 0$, for every compact  set  $K$ of  $\R$, there
exists $C>0$   such that :
$$h^n I_h^{m, tr }  (Op_h^{weyl} (R_N (., t_N , ... , t_1 , h)) ) \leq
C_{m K}$$
if $0< h \leq 1$, $(t_1, ... t_N) \in \Delta_N (0, t)$, and $t\in
K$. Hence by  lemma 4.4,   we deduce that the  iterated commutators
of $ G_h (t , t_N) (Op_h^{weyl} (R_N (., t_N , ... , t_1 , h)) )$
with the position and momentum  operators are themselves trace
class, and   that  there  exists
  another constant $C_{m K}$ such that:
$$h^n I_h^{m, tr }  \Big ( G_h (t , t_N) (Op_h^{weyl} (R_N (., t_N , ... , t_1 , h)) )
\Big )  \leq C_{m K}. $$
We can therefore  write, if  $A_h = Op_h^{weyl}(F)$, for every
multi-index  $ (\alpha , \beta)$ and for every integer $N$:
$$ h^n \Vert (ad \
Q(h))^{\beta }(ad \ P(h))^{\alpha } E_N (t , h)  \Vert _{{\cal
L}^1({\cal H})} \leq C_{\alpha \beta N }$$
By reporting  this in  (4.9), and by  choosing  $N
=|\alpha|+|\beta|$, one deduces   (4.18).  Using  the
characterization of C. Rondeaux (proposition 2.1), the  theorem 4.2
follows.

\hfill \carre

\vskip 8cm

{\it  Proof  of  theorem  1.1.}

\bigskip

 Let  $\rho_h (t)$  be a  classical  solution  of (TDHF) verifying   the hypotheses
 of  theorem 1.1.  Let us denote by  $V_h (t)$
the  operator of  multiplication by the following function
$$ x \rightarrow  V_q ( x, \rho_h (t) ) =
 V(x) + Tr ( W _x \circ \rho_h (t) ),  \hskip 1cm W_x (y) = W (x-y). \leqno (4.19)$$
Under  the  hypotheses of  theorem  1.1,  $\rho_h (0) \geq 0$. By
the results of [BdPF1] and [BdPF2],  $\rho_h (t)\geq 0$ for all $t$,
and the trace norm of  $\rho_h (t)$ is constant.  Since all the
derivatives of $V$ et $W$ are bounded , one  has:
$$ |\partial_x^{\alpha}V_q ( \rho_h (t) )(x) |  \leq C_{\alpha},
\hskip 2cm   (x , t)\in \R^n\times \R.  \leqno (4.20) $$
Let   $\widehat H_h(t)$   denote   the  operator defined  in  (4.3),
where $V(t)$ is  the multiplication by  $V_q ( x, \rho_h (t) )$.
With these notations,  the  equation (TDHF)  has  the form:
$$ih {d \rho_h(t) \over dt} =
  [\widehat H_h (t) , \rho_h(t) ].   \leqno (4.21)$$
We note that  $V(t)$ depends on $h$. However in theorem 4.2, $V(t)$
may depend  on a parameter  that could  be  $h$. The only
requirement  is that $V_q ( ., \rho_h (t) )$  be bounded in
$W^{\infty \infty}(\R^n)$ independently  of $h$, which is the case.
By denoting  $G_h (t , s)$  the unitary  propagator associated  to
the Hamiltonian $ H_h (t)$ as in proposition 4.1, one has therefore:
$$ \rho_h (t) = G_h (t , 0) ( \rho_h (0))=
 G_h (t , 0) (  Op_h^{weyl } (F_h)). $$

Theorem 1.1  is therefore a  particular case  of theorem  4.2.

\bigskip

{\bf 5. Proof  of  theorem 1.2.}

\bigskip

We  are going to  state precisely  the  "explicit"  construction  of
an "approximate  solution to order  $N$", denoted by $\rho_h^{(N)}
(t)$, of the  (TDHF) equation.  The exact solution  $\rho_h(t)$ is
determined by the interaction  potentials  $V$ et $W$, which belong
to  $W^{\infty \infty }(\R^n)$,  and  the  initial  data
$\rho_h(0)$.  Under  the  hypotheses of  theorem 1.1, the operator
$\rho_h(0)$ is $\geq 0$, and  is of the  form
$$ \rho_h (0) = (2\pi h)^n\  Op_h^{weyl}(F_h)$$
with  $F_h$ in  $W^{\infty 1}(\R^{2n})$. This function may
depend on  $h$, but  stays bounded  in $W^{\infty 1}(\R^{2n})$
independently  of  $h$ in  $(0, 1]$.

\bigskip

We look for an approximate solution  with the ansatz (1.13), where
$F^{(N)} (t , h)$ is a function $\R^{2n}$ of the form (1.11),  the
functions $u_j(., t)$  being in $W^{\infty 1} (\R^{2n})$.  We
associate to  this approximate solution  the average quantum
potential:
$$ V_q ( x , \rho_h^{(N)} (t))  = V(x) + Tr ( W_x \rho_h^{(N)} (t)). \leqno (5.1)$$
One knows  that, if  $F$ is in $W^{\infty 1}(\R^{2n})$ and  $G$  in
$W^{\infty \infty }(\R^{2n})$, one  has:
$$ Tr ( Op_h^{weyl}(F) \circ Op_h^{weyl}(G)) = (2\pi h)^{-n} \int
_{\R^{2n}} F(X) G(X) dX. $$
Therefore, if  $\rho_h^{(N)} (t)$ is defined by  (1.13),  we have:
$$ V_q ( x , \rho_h^{(N)} (t))  = V_{cl} ( x ,  F^{(N)} (., t , h) ), \leqno (5.2) $$
where, for every function  $G$ in  $L^1(\R^{2n})$,  the function $
V_{cl}  ( . , G)$ is  defined as in  (1.1):
$$ V_{cl}  ( x , G)  = V(x) + \int _{\R^{2n}} W ( x-y) G(y , \eta) dy
d\eta.  \leqno (5.3)$$
One shows similarly  that:
$$ V_q ( x , \rho_h (t))  = V_{cl} ( x,  u_h (. , t) ),  \leqno (5.4) $$
where   $u_h$ is defined in (1.7).  For all   suitable  functions
$A$ and $B$ , let us denote by  $M_h (A , B)$ the Moyal  bracket of
$A$ and $B$, defined  in  (3.2). With these notations, the function
$u_h(., t)$ defined  in  (1.7) must  verify:
$$ {\partial u_h (., t)  \over \partial t}  + 2 \sum _{j= 1}^n
\xi_j {\partial u_h(., t)\over \partial \xi_j} = {1\over ih} M_h
\Big ( V_{cl} ( u_h (., t)), u_h(., t) \Big ).  \leqno (5.5)$$
For all  functions $A$ and  $B$ in  $C^{\infty}(\R^{2n})$, and for
every  integer  $k\geq 0$, let  $C_k(A , B , .)$   the function
defined in  (3.4). We set $C_0 (A , B  )  = 0$. One has  $C_1(A , B)
= {1\over i} \{ A , B \}$. We  will choose  the  $u_j$ in  a such a
way  that the equation (5.5)   is  approximatively  verified.

\bigskip

The  construction of the  functions $u_j$   of  theorem 1.2 is
detailed  in the following proposition.

\bigskip

{\bf Proposition 5.1.} {\it There  exists a sequence of  functions
$(X , t) \rightarrow u_j (X , t)$ on  $\R^{2n} \times \R$ ($j\geq
0$), such that:
\smallskip

a) The function  $t \rightarrow u_j (., t, h)$ is  $C^{\infty}$ on
$\R$ valued  in $W^{\infty, 1}(\R^{2n})$, bounded in that
space  independently  of  $h$  in  $(0, 1]$ and of  $t$ in every
compact set  of $\R$.

\smallskip

b)  One   has:
$$ u_0 (X , 0) = F_h(X),  \hskip 2cm u_j (X , 0, h) = 0, \ \ \ \ \  (j\geq 1). $$

  \smallskip

c)  For  every  $N$, the  function $u_N (X , t, h) $ verifies:
$${\partial u_N  \over \partial t} + 2 \sum _{j=1}^n \xi_j {\partial
u_N  \over \partial x_j}  = {1\over i}  \sum _{j+ k + \ell = N+1}
C_k \Big ( V_{cl} ( ., u_j  (., t, h)) \ , \ u_{\ell} (., t, h) \Big
).  \leqno (5.6) $$

}

\bigskip

In the sum (5.6)  the  indices $j$ et $\ell $  are $\geq 0$ and $k$
is  $\geq 1$.

\bigskip

{\it D\'etermination  of  $u_0$.} For $N=0$, the equation (5.6)
reduces  to   the Vlasov's  equation:
$${\partial u_0  \over \partial t} + 2 \sum _{j=1}^n \xi_j {\partial
u_0  \over \partial x_j} = \sum _{j= 1}^n {\partial \over \partial
x_j} V_{cl} (u_0 (., t)) \  {\partial u_0 (., t)\over \partial
\xi_j}$$
It is well known  that there  exists a unique solution of the Cauchy
problem, verifying  this equation and  $u_0 (. , 0, h)= F_h$, where
$F_h$ is  a  given  function in  $W^{\infty 1}(\R^{2n})$. One knows
that then  the  function $u_0$ is continous on $\R$ valued in
$W^{\infty 1}(\R^{2n})$. If  $F_h$ stays bounded in $W^{\infty
1}(\R^{2n})$ independently  of  $h$, it is also the case for $u_0(.,
t , h)$.

\medskip

{\it D\'etermination  of $u_N$ ($N\geq 1$).}  For every  $N \geq 1$,
the equation (5.6) can be written as:
$${\partial u_N  \over \partial t} + 2 \sum _{j=1}^n \xi_j {\partial
u_N  \over \partial x_j} = \sum _{j= 1}^n {\partial \over \partial
x_j} V_{cl} (u_0 (., t, h)) \  {\partial u_N (., t)\over \partial
\xi_j} + \sum _{j= 1}^n {\partial \over \partial x_j} V_{cl} (u_N
(., t, h)) \ {\partial u_0 (., t, h)\over \partial \xi_j} + G_N ( X
, t, h),$$

$$ G_N = {1\over i}  \sum _{j+ k + \ell = N+1 \atop j < N, \ell < N}
C_k \Big ( V_{cl} ( u_j  (., t, h)) \ , \ u_{\ell} (., t, h) \Big ).
$$
One also  requires  that   $u_N (X , 0, h)=0$. To solve  this
equation,  by dropping  the parameter   $h$ for the sake of
simplifying notations, let us denote by $ X \rightarrow \varphi_t(X)
= (q(t , X) , p(t , X))$ the the Hamiltonian flow, solution of:
$$ q'(t, X) = 2 p(t, X) \hskip 2cm p'(t, X) = - \nabla V (  q(t, X)) - \int
_{\R^{2n}} \nabla W ( q(t, X) - y) u_0 ( y , \eta, t)dy d\eta, $$
verifying :
$$q(0, X) =x \hskip 2cm p(0, X) = \xi \hskip 2cm X = (x , \xi). $$
The function $v_N$
defined  by $v_N (X , t) = u_N ( \varphi_t (X), t)$ must
verify:
$${\partial v_N  \over \partial t}  = \sum _{j= 1}^n {\partial u_0 \over \partial \xi_j}
(\varphi_t (X) , t) \int _{\R^{2n}} {\partial W \over \partial x_j}
( q_t (X) - y)\ u_N ( y , \eta , t) \ dy d\eta + \widetilde G_N (X ,
t), $$
$$ \widetilde G_N (X , t) = G_N ( \varphi_t(X), t). $$
By using in  the integral the  change of variable $(y , \eta)
= \varphi_t(z , \zeta )$, whose  jacobian equals $1$, we  see  that
$v_N$ must satisfy
$${\partial v_N  \over \partial t}(X , t)   = \widetilde G_N (X , t)  +
\int _{\R^{2n}} A (X , Y , t) v_N (Y , t) dY, $$
$$ A(X , Y , t) = \sum _{j= 1}^n {\partial u_0 \over \partial \xi_j}
\big (\varphi_t (X) , t \big )\   {\partial W \over \partial x_j}
\big ( q_t (X) - q_t (Y) \big  ).$$
Moreover, one must have  $v_N (., 0) = 0$. According to standard
results on  the Vlasov  equation, one  knows that $\nabla u_0 (.,
t)$ is  in  $W^{\infty 1}(\R^{2n})$, bounded  when $t$ varies  in a
compact  set.  The same  is true for $\nabla \varphi_t$. If the
$u_j$ ($0 \leq j < N$) have been built  with the  properties of
proposition 5.1, one sees that $G_N (., t)$ is in $W^{\infty
1}(\R^{2n})$, bounded  when $t$ varies  in  a  compact set. It is
also the case for $\widetilde G_N (., t)$.  The resolution of the
Cauchy problem verified  by $v_N$  and the one verified by $u_N$  is
standard.

\hfill \carre

\bigskip

To prove theorem 1.2, we will show that the functions $F^{(N)}(t ,
h) (X)$ defined  in (1.11), starting from the $u_j(., t , h)$  of
proposition 5.1, and  the operators $\rho_h^{(N)}(t)$ defined  in
(1.13),  verify (1.12)  and (1.14). The next proposition is an
intermediate step.

\bigskip

{\bf Proposition 5.2.} {\it Let  $\rho_h(t)$ be an exact solution of
(TDHF) verifying the hypotheses of theorems 1.1   and  1.2.
Let  $\widehat H_h(t)$  be  the  operator defined  in (4.3),  where $V(t)$
is  the  multiplication by  $V_q ( x, \rho_h (t) )$. Then the  above constructed approximate solution
$\rho_h^{(N)} (t)$  verify:
$$ ih {d \rho_h^{(N)} (t) \over dt} =
  [ \widehat H_h(t)  , \rho_h^{(N)} (t) ]  +
  Op_h^{weyl } \Big ( S ^{(N)}_h (., t ) \Big )  \leqno (5.7) $$
where  $ S ^{(N)}_h (., t ) $ is in  $W^{\infty 1} (\R^{2n})$ and
verify, for every integer  $\alpha $:
$$ \Vert \nabla ^{\alpha } S^{(N)}_h  (., t  )\Vert _{ L^1(\R^{2n})}
\leq C_{\alpha  N } (t) \Big [ h^{N+2}  + h \Vert \rho_h(t) -
\rho_h^{(N)}(t) \Vert _{{\cal L}^1({\cal H})}\Big ] \leqno (5.8)$$
where $C_{\alpha  N } (t)$  is a bounded function on every compact set.

}

\bigskip

{\it Proof.} If the  $u_j$ are  determined  relying on proposition
5.1, the function $F^{(N)}$ defined  in (1.11) verifies:
$$ {\partial F^{(N)}  \over \partial t} + 2 \sum _{j=1}^n \xi_j {\partial
F^{(N)}   \over \partial x_j}  = {1\over h}\sum _{k= 1}^{N+1} h^k
C_k \Big ( V_{cl} (.,  F^{(N)} (., t, h)) \ , \ F^{(N)} (., t, h)
\Big ) + \Phi ^{(N)} (. , t, h), \leqno (5.9)$$
where  $\Phi ^{(N)} (. , t, h)$ is a function in  $W^{\infty
1}(\R^{2n})$,  such that:
$$\Vert \nabla^{\alpha}  \Phi ^{(N)} (. , t, h)
\Vert _{L^1 (\R^{2n})} \leq  h^{N+1} C_{\alpha  N}(t).  \leqno
(5.10)
$$
We  define an approximate version of the operator
 $\widehat H_h (t)$ by setting :
$$\widehat H_h^{APP}  (t) = - h^2 \Delta + V_{cl} ( F^{(N)}
(., t, h)).  \leqno (5.11) $$
Since  $F^{(N)}$ verifies (5.9),  we may write:
$$ih {d \rho_h^{(N)} (t) \over dt} =
  [\widehat H_h^{APP}  (t) , \rho_h^{(N)} (t) ]  +
  Op_h^{weyl} \Big (  T _h^{(N)} (., t ) \Big ),  \leqno (5.12) $$
where  the  function  $T_h^{(N)}(., t)$  is  defined  by:
$$   T^{(N)}_h (., t) =  h   \Phi ^{(N)} (. , t, h)+   R_h^{(N+2)}
\big  ( V_{cl} (.,  F^{(N)}  (., t, h)) , F^{(N)}  (., t, h) \big ).
$$
(For all functions $A$ et $B$ verifying  the  hypotheses of theorem
3.1, we denote by $R_h^{(N)} (A , B, .)$  the  function
 associated by  theorem  3.1.  to such functions.)  Then by the
definition (5.3)  of the map $V_{cl}$, and point a) of proposition
5.1, we can write:
$$ \Vert \nabla_x^{\alpha} V_{cl}  (.,  F^{(N)}  (., t, h)) \Vert _{L^{\infty}(\R^{2n})}
 \leq C_{\alpha  N}(t),
\hskip 2cm \Vert \nabla ^{\beta}  F^{(N)}  (., t, h)) \Vert _{L^{1
}(\R^{2n})}  \leq C_{\beta  N}(t).  \leqno (5.13)$$
Using these upper bounds  and following  theorem 3.1   on the
 Moyal  bracket, we may write:
$$ \Big \Vert \nabla^{\ell}   R_h^{(N+2)}
( V_{cl} \big ( ., F^{(N)}  (., t, h)) , F^{(N)}  (., t, h)   \big )
\Big \Vert _{L^1 (\R^{2n})}  \leq C_{\ell N}(t) h^{N+2}.
$$
Acccording to these upper bound estimates,  and the estimates (5.10)
of the derivatives  of  $\Phi ^{(N)} (. , t, h)$, one  has:
$$ \Vert  \nabla^{\alpha}  T ^{(N)}_h  (., t) \Vert _{L^1 (\R^{2n})}
\leq C_{\alpha  N} (t) h^{N+2}.  \leqno (5.14) $$
According  to  (5.12)  and  since
$$\widehat H_h^{APP}  (t) - \widehat H_h  (t) =
V_q (.,  \rho_h(t) ) - V_q (.,  \rho_h^{(N)} (t)),
$$
we write  the    equality (5.7)  with:
$$ S^{(N)}_h  (., t) =  T^{(N)} _h (., t) +
M_h \Big (    V_q (.,  \rho_h(t))  -  V_q (.,  \rho_h^{(N)} (t))  ,
F^{(N)} (., t , h) \Big ).  \leqno (5.15) $$
One has:
 $$\Big \Vert \nabla _x^{\alpha} \Big (  V_q (.,  \rho_h(t))  -  V_q (.,  \rho_h^{(N)}
 (t))\Big )
 \Big \Vert _{L^{\infty} (\R^{2n})} \leq
C_{\alpha } \     \Vert  \rho_h(t) -  \rho_h^{(N)}(t) \Vert _{{\cal
L}^1({\cal H})}.  $$
Using all of these estimates  and the $L^1$  norm estimates  (5.13)
of  $F^{(N)}(., t, h)$  and by using  theorem 3.1  on the
 Moyal  bracket,  it results that:
$$ \Big  \Vert \nabla^{\alpha} M_h \Big (   V_q ( \rho_h(t))  - V_q ( \rho_h^{(N)} (t))  ,
 F^{(N)}(., t , h)
\Big ) \Big  \Vert _{ L^1(\R^{2n})} \leq C (t) h \Vert  \rho_h(t) -
\rho_h^{(N)}(t)
  \Vert _{{\cal L}^1({\cal H})}. \leqno (5.16)$$
The norm upper bound estimate 5.8) of  $ S^{(N)}_h  (., t )$ results from
(5.15), (5.14)  and  (5.16).

\bigskip

{\it End of the proof of  theorem 1.2.}  Let  $U_h (t , s)$ and $G_h
(t , s)$ be the unitary propagator and  the mapping, defined in
proposition 4.1,  associated to  the operator
  $\widehat H_h(t)$ of proposition  5.2.  The  comparison of  equalities
(4.21) (verified by the exact solution) and  (5.7) (verified by the
approximate solution ),  and the  Duhamel principle,  allows us  to
write:
$$ \rho_h(t) -  \rho_h^{(N)}(t) = {i \over h} \int _0^t G_h (t ,
s) \Big ( Op_h^{weyl} ( (S_h^{(N)}  (., s) )) \Big )  ds.  \leqno
(5.17)$$
Since $U_h(t , s)$  is  unitary, the map  $G_h (t , s)$ conserves
the trace norm, and from this  we may deduce that:
$$\Vert  \rho_h(t) -  \rho_h^{(N)}(t) \Vert _{{\cal L}^1({\cal H})}
\leq {1\over h} \int _0^t \Vert Op_h ^{weyl}  ( S_h^{(N)}  (., s))
\Vert _{{\cal L}^1({\cal H})}  \ ds. $$
Using the upper bounds   (5.8) of  Proposition 5.2,  and according
to  proposition  2.1,   we  obtain:
$$ \Vert  \rho_h(t) -
\rho_h^{(N)}(t) \Vert _{{\cal L}^1({\cal H})} \leq {1\over h}  \int
_0^t C(s)  [ h^{N+2}  + h \Vert  \rho_h(s) - \rho_h^{(N)}(s) \Vert
_{{\cal L}^1({\cal H})}] \  ds. $$
By the Gronwall  lemma, we deduce, with a different constant that
$$\Vert  \rho_h(t) -  \rho_h^{(N)}(t) \Vert _{{\cal L}^1({\cal H})}
\leq C(t) h^{N+1}.$$
Therefore point (1.14) of  theorem 1.2 is  proved. We deduce from
this inequality and from  (5.8) that:
$$ \Vert \nabla ^{\alpha } S^{(N)} _h (. , t )\Vert _{ L^1(\R^{2n})}
\leq C_{\alpha  N } (t)  h^{N+2}.
$$
where  $C_{\alpha  N } (t)$  is a bounded function on every compact
set. From  proposition 2.1 and lemma 4.4,  for every multi-index
$(\alpha \beta)$,  the operators:
$$h^{-N-2}\ (ad \ Q(h))^{\beta }(ad \ P(h) )^{\alpha }G_h(t , s) ( Op_h^{weyl} \Big (
S^{(N)}_h (., s  )\Big )$$
are trace class, and their trace norm  is  independently  bounded of
 $(t , s)$ in a  compact set of  $\R$, and of  $h$ in  $(0, 1]$.
By  (5.17), for every multi-index  $(\alpha , \beta)$,  and for
every  $N>0$, there exists  a function $C_{\alpha N }(t)>0$, bounded
on every   compact  set   of $\R$, such that:
$$ \Vert (ad \ Q(h))^{\beta }(ad \ P(h) )^{\alpha }
\Big ( \rho_h(t) -  \rho_h^{(N)}(t)\Big )
 \Vert _{{\cal L}^1({\cal H})} \leq C_{\alpha N}(t)  h^{N+1}. $$
Using  proposition 2.1, point b), we  deduce:
$$ (2 \pi h )^{-n} \Vert \sigma_h^{weyl} \Big ( \rho_h(t) -  \rho_h^{(N)}(t)\Big )
\Vert _{L^1(\R^{2n})} \leq C(t) h^{(N+1) -{2n+2\over 2}}.$$
In other words, with the  notations of  theorem 1.2:
$$ \Vert u_h(., t) - F^{(N)} (., t , h) \Big )
\Vert _{L^1(\R^{2n})} \leq C(t) h^N. $$
Theorem  1.2  results  from the above considerations.

\bigskip

{\bf Appendix A. Proof of proposition 2.1.}

\bigskip

The proof of  proposition  2.1  calls upon a different notion of
symbol. One can  associate   to every bounded   operator $A$ in
${\cal H}$ a  $S_h(A)$ sur $\R^{2n} \times \R^{2n}$
defined by:
$$ S_h (A) (X , Y ) = { < A \Psi_{Xh} , \Psi _{Y h}> \over <  \Psi_{Xh} , \Psi
_{Y h}>}, \leqno (A.1) $$
where the  $\Psi_{Xh}$ are defined in  (2.14). An explicit
computation of integrals  shows that:
$$ | < \Psi_{Xh} , \Psi_{Yh}>| = e^{- {1 \over 4h} |X-Y|^2} \hskip  3cm \Vert \Psi_{Xh}\Vert =
1. \leqno (A.2)$$
Consequently:
$$
  \Big |
S_h (A) (X , Y ) \Big |  =  e^{{1\over 4h} |X-Y|^2} |< A \Psi_{Xh} ,
\Psi_{Y h} > |. \leqno (A.3) $$

\bigskip

The interest  of considering this  function $S_h(A)$   (which is up to a slight
modification what  Folland [F]  calls Wick  symbol), is
that it is related to the Weyl symbol, in both ways,  by integral
operators   (proposition A1), and that it can be majorized (in one norm)
and  minorized  (in a different norm)  by the trace norm  of $A$ (proposition A2).

\bigskip

{\bf Proposition A.1.} {\it  The Weyl symbol   of an operator $A$
is  related to the function $S_h (A)$  by
$$ S_h (A) (X , Y) = (\pi h)^{-n} \int _{\R^{2n}}  e^{-{1 \over
h} (Z-X).(\overline Z -  \overline Y)}\ \sigma_h^{weyl}(A) (Z) \ dZ,
\leqno (A.4)$$
$$\sigma_h^{weyl} (A) (Z) = 2^n  ( 2\pi h)^{-2n} \int _{\R^{4n}}
S_h (A) (X , Y) \ K_h (X , Y, Z) \ dXdY,  \leqno (A.5) $$
$$K_h (X , Y, Z)  = e^{ -{1\over h} (\overline Z - \overline X) ( Z
-Y) - {1\over 2h} |X-Y|^2}. \leqno (A.6)$$
}

\bigskip

By  the definition  formula (2.3)  of the Weyl calculus, one  has:
$$ A = (\pi h)^{-n} \int _{\R^{2n}} \ \Sigma _{Zh} \  \sigma_h^{weyl}(A) (Z) \
dZ, $$
where  $\Sigma _{Zh}$  is the operator defined  in (2.4).  An
explicit   computation shows that:
$$ { < \Sigma _{Zh} \Psi_{X h} , \Psi _{Y h} > \over
<  \Psi_{X h} , \Psi _{Y h} > } = e^{-{1 \over h} (Z-X).(\overline Z
-  \overline Y)}. \leqno (A.7) $$
The equality  (A.4)  follows.  By  the fundamental   formula (2.16)
of  coherent states, on has:
$$ A = (2 \pi h)^{-2n} \int _{\R^{4n}}  <A  \Psi_{X h}, \Psi _{Y h} > P_{X Y h}
dX dY, \leqno (A.8) $$
where  $P_{X Y h}$  is the operator defined by:
$$ P_{X Y h} f = <  f , \Psi_{X h} > \Psi_{Yh}. \leqno (A.9) $$
One knows from  (2.5)  that:
$$ \sigma_h ^{weyl} ( P_{X Y h }) (Z) = 2^n {\rm Tr } ( P_{X Y h }
\circ \Sigma_Z) = 2^n <  \Sigma_{Zh} \Psi_{Y h } , \Psi_{Xh} >.  $$
By computation (A.7) (where $X$ et $Y$  are permuted ) we may
deduce:
$$ \sigma_h ^{weyl} (  A) (Z) = 2^n  ( 2\pi h)^{-2n} \int _{\R^{4n}}
S_h (A) (X , Y)\ |< \Psi_{Xh} , \Psi_{Y h }> |^2 e^{ -{1\over h}
(\overline Z - \overline X) ( Z -Y)}  dX dY. $$
Using the  equality (2.15)  on the  scalar  product  of coherent
states, we  obtain (A.5).

\bigskip

{\bf Proposition A.2.} {\it Let  $A$ be a trace class operator and
$G$ a function in $L^1(\R^{2n})$. Then one has :
$$ (2\pi h)^{-2n}  \int _{\R^{4n}}    \Big  |< A \Psi_{X h}
, \Psi_{Y  h} >   G\left ( {X-Y \over \sqrt h} \right )  \Big |  \
dX dY \leq (2\pi)^{-n}  \Vert G \Vert _{L^1(\R^{2n})} \Vert A \Vert
_{{\cal L}^1({\cal H})}, \leqno (A.10) $$
$$ \Vert A \Vert
_{{\cal L}^1({\cal H})} \leq (2 \pi h)^{-2n}
 \int _{\R^{4n}} \Big  |< A \Psi_{X h}
, \Psi_{Y  h} > \Big | dX dY. \leqno (A.11) $$

 }

\bigskip

{\it Proof.} We may write $A = B_1 B_2$,  where $B_1$ and  $B_2$
are  Hilbert-Schmidt. By using the fundamental  property
 (2.16) of coherent  states one sees  that, for all
$X$ et $Y$ in $R^{2n}$:
$$ < A \Psi_{Xh} , \Psi_{Y h} > = < B_2 \Psi_{Xh} , B_1^{\star}  \Psi_{Y h} >
= (2 \pi h)^{-n} \int _{\R^ {2n}} u_{Zh} (X)  \  v_{Zh} (Y) \  dZ,$$
where we have denoted by  $u_{Zh } (X) = < B_2 \Psi_{Xh} ,
\Psi_{Zh}> $ and $v_{Zh } (X) = < \Psi_{Zh} ,  B_1^{\star} \Psi_{X
h} > $. Let $I_h$ be the left hand side  of  (A.10). By  the Schur
lemma:
$$ I_h \leq (2\pi h)^{-3n} h^n  \Vert G \Vert _{L^1(\R^{2n})} \  \int _{\R^{2n}} \Vert u_{Zh} \Vert
 _{L^2(\R^{2n})} \ \Vert v_{Zh} \Vert _{L^2(\R^{2n})} dZ. $$
 By (2.16), we have $\Vert u_{Zh} \Vert
 _{L^2(\R^{2n})}  = (2 \pi h)^{n/2} \Vert B_2^{\star} \Psi_{Zh} \Vert  $ et $ \Vert v_{Zh} \Vert
 _{L^2(\R^{2n})}  = (2 \pi h)^{n/2}  \Vert B_1 \Psi_{Zh} \Vert$.
Hence :
$$ I_h \leq  (2\pi h)^{-2n} h^n  \Vert G \Vert _{L^1(\R^{2n})} \    \int _{\R^{2n}} \Vert
B_1 \Psi_{Zh} \Vert \ \Vert B_2^{\star} \Psi_{Zh} \Vert \ dZ$$
By  the fundamental  property   of coherent  states:
$$(2\pi h)^{-n} \int _{\R^{2n}}  \Vert
B_j \Psi_{Zh} \Vert ^2 dZ = (2\pi h)^{-n} \int _{\R^{2n}}  < B_j
^{\star} B_j  \Psi_{Z h} , \Psi_{Zh} > \ dZ$$
$$ = {\rm Tr } ( B_j ^{\star } B_j) = \Vert B_j\Vert _{{\cal L}^2 ({\cal H})}^2$$
where  $ \Vert B_j\Vert _{{\cal L}^2 ({\cal H})}$ is the
Hilbert-Schmidt norm  of $B_j$. Therefore:
$$ I_h  \leq   (2\pi)^{-n}   \Vert G \Vert _{L^1(\R^{2n})} \ \Vert B_1 \Vert _{{\cal
L}^2 ({\cal H})}\ \Vert B_2 \Vert _{{\cal L}^2 ({\cal H})}. $$
By taking  the infimum  over all   the decompositions
 $A = B_1 B_2$,  one gets  (A.10).  The inequality  (A.11)  is then deduced
from the equality (A.8)  since the  operators $P_{X Y h}$ have
a trace norm  equal  to  $1$.

 \hfill \carre

\bigskip

{\it Proof  of proposition 2.1.} For point  a),  we use the equality
(A.4), by integrating  by parts , as in Rondeaux [R]. Hence we have
shown:
$$ (2 \pi h)^{-2n}
 \int _{\R^{4n}}  e^{-{1\over 4h}|X-Y|^2}
\ |S_h(A) (X , Y )| dX dY \leq C h^{-n} \sum _{|\alpha |+|\beta|\leq
2n+2} h^{(|\alpha |+|\beta |)/2} \Vert \partial _x^{\alpha} \partial
_{\xi}^{\beta}F \Vert _{L^1(\R^{2n})} . $$
One deduces then  the upper bound estimate  of the  trace  norm  of  $A$ (point a) by
using the  second point  of proposition A2.

\smallskip

 For  points b) et c), we are going to integrate by parts in the second equality (A.5)
 of  proposition A.1. One verifies  that  the  function  $K_h$ defined in  (A.6) is invariant
 by the following  differential  operator:
$$ L(h) K_h = K_h \hskip 2cm L(h) = \left ( 1 + {|X-Y|^2 \over h}
\right )^{-1} \ ( 1 +  ( \overline Y - \overline  X ) \partial _X ).
$$
Thus, equality  (A.5)  implies, for every integer $N$:
$$| \sigma_h^{weyl} (A) (Z)| \leq  2^n  ( 2\pi h)^{-2n} \int
_{\R^{4n}} \ |K_h (X , Y, Z)| \ |(^t L(h))^N S_h (A) (X , Y) |\ dX
dY .$$
One  verifies that:
$$ |K_h (X , Y, Z)| =e^{ -{1\over h}  |Z - {X+Y\over 2}|^2 - {1\over
4h} |X-Y|^2} . $$
On chooses $N = 2n+2$. There exists  $C>0$  such that:
$$  | \sigma_h^{weyl} (A) (Z)| \  \leq ... \hskip 9cm \ $$
$$...  \leq  C \sum _{|\alpha | \leq 2n+2}
 h^{|\alpha|/2} ( 2\pi h)^{-2n} \int _{\R^{4n}} |K_h (X , Y, Z)|  G
\left ( { X-Y\over \sqrt h } \right )\ |\partial _X^{\alpha}S_h (A)
(X , Y) |\ dX dY, $$
$$ G(X) = ( 1 + |X|)^{-2n-2}. $$
By the formula on the commutators:
$$  | \sigma_h^{weyl} (A) (Z)| \  \leq ... \hskip 9cm \ $$
$$ \leq  C \sum _{|\alpha |+|\beta| \leq 2n+2} h^{-2n -(|\alpha|+|\beta|)/2}
 \int _{\R^{4n}}e^{ -{1\over h}  |Z - {X+Y\over 2}|^2 - {1\over
4h} |X-Y|^2}   G \left ( { X-Y\over \sqrt h } \right )\  |S_h (A
_{\alpha \beta h}  ) (X , Y)| \ dXdY, $$
$$ A _{\alpha \beta h} = (ad\  P(h) )^{\alpha} (ad \
Q(h))^{\beta } A.  $$
The above equality can be also written  as:
$$  | \sigma_h^{weyl} (A) (Z)| \  \leq ... \hskip 9cm \ $$
$$ \leq  C \sum _{|\alpha |+|\beta| \leq 2n+2} h^{-2n -(|\alpha|+|\beta|)/2}
 \int _{\R^{4n}}e^{ -{1\over h}  |Z - {X+Y\over 2}|^2}   G \left ( { X-Y\over \sqrt h } \right )\
 |< (A _{\alpha \beta h} \Psi_{Xh}, \Psi_{Yh} > | \ dXdY. $$
Inequality (2.13) is a consequence  of the first point  of
proposition A.2, and point b) follows easily.

\bigskip

{\bf Appendix  B.  Proof   of theorems 3.1 and  3.2. }

\bigskip

{\it First step, common  to  both theorems 3.1 and 3.2.}  We know
that, for all suitable functions $F$ et $G$ , $Op_h^{weyl}(F) \circ
Op_h^{weyl}(G) = Op_h ^{weyl}(C_h(F , G , .)$, with:
$$ C_h(F , G , X) = (\pi h)^{-2n}\int _{\R^{4n}} e^{- {2i \over h} \sigma (Y
-X , Z- X)} F(Y) G(Z) dY dZ$$
where $\sigma $ is the  symplectic  form  $\sigma ((x , \xi), (y ,
\eta)) = y \xi  - x \eta$. Consequently the  Moyal bracket
$M_h(F , G, .)$ is  defined  by $ M_h(F , G , X) = C_h (F , G , X)
- C_h (G , F , X)$.  Thus, it suffices  to write an asymptotic  expansion   $C_h (F , G , .)$.
We may write
$C_h (F , G , X) = \Phi_h (X , 1) $, by setting, for every $\theta
\in [0, 1]$:
$$ \Phi_h (X , \theta) = (\pi h)^{-2n}\int _{\R^{4n}} e^{- {2i \over h} \sigma (Y
-X , Z- X)}\  F(Y) G( X + \theta(Z -X)) dYdZ.  $$
Consequently, for every integer  $N$:
$$C_h (F , G ,  X) = \sum _{k=0}^{N-1} { 1 \over k!} \partial_t^k
\Phi_h (X , 0)  + \widetilde R_h^{(N)} (F , G , X) $$
with:
$$ \widetilde R_h ^{(N)} (F , G , X) = \int_0^1 {(1-\theta )^{N-1} \over (N-1)!} \partial
_{\theta}^N \Phi_h ( X , \theta ) d\theta .    $$
One sees, using integration by parts, that:
$$\partial_{\theta}^k  \Phi (X , \theta , h) = \left ( {h\over 2i} \right )^k
 (\pi h)^{-2n}\int _{\R^{4n}} e^{- {2i \over h} \sigma (Y
-X , Z- X)}\  \Big ( \sigma (\nabla_1 , \nabla _2)^k ( F \otimes G)
\Big ) ( Y , X + \theta  (Z -X)) \ dY dZ . $$
If a function $\Phi$ depends  only on the  $Y$ variable   one has (in the sense of
distributions):
$$(\pi h)^{-2n}\int _{\R^{4n}} e^{- {2i \over h} \sigma (Y -X , Z-
X)} \Phi (Y)  dY dZ = \Phi (X)$$
and  similarly  if $\Phi$ depends only on the $Z$  variable. For
$\theta =0$, one has, by the  above two  equalities:
$$\partial_{\theta}^k  \Phi (X , 0 , h) =  \left ( {h\over 2i} \right )^k
  \sigma (\nabla_1 , \nabla _2)^k ( F \otimes
G)  (X , X).  $$
Therefore we  do have indeed  the equality  (3.3)  of  theorem 3.1,
by setting:
$$ R_h^{(N)} (F , G , X) = \widetilde R_h^{(N)} (F , G , X) -  \widetilde R_h^{(N)} (G , F ,
X).  \leqno (B.1)$$
It remains to obtain  an upper bound  for the norm of the two above
terms.   One also has:
$$ \widetilde R_h ^{(N)} (F , G ,X) = \left ( {h\over 2i} \right )^N
(\pi h)^{-2n}\int _{\R^{4n}\times [0, 1]} {(1-\theta )^{N-1} \over
(N-1)!} K_h (X , Y , Z) \Psi  (X , Y , Z , \theta ) dY dZ d\theta
$$
where
$$K_h(X , Y , Z) =  e^{- {2i \over h} \sigma (Y -X , Z- X)}$$
$$ \Psi (X , Y , Z, \theta  ) =  \Big ( \sigma (\nabla_1 , \nabla _2)^N ( F \otimes G)
\Big ) ( Y , X + \theta  (Z -X)).  $$
The function  $K_h$   is invariant by the following  operator:
$$ L_h = \left ( 1  + 4 {|X-Y|^2 \over h} + 4 {|X-Z|^2 \over h}     \right )^{-1}
(1 - h \Delta _Y - h \Delta _Z )$$
Therefore, for all integers  $K$ and  $\ell$:
$$ |\nabla ^{\ell} \widetilde R_h ^{(N)} (F , G , X)| \leq \left ( {h\over 2} \right )^N (\pi h)^{-2n}\int _{\R^{4n}\times [0, 1]}
{(1-\theta )^{N-1} \over (N-1)!}  |\nabla^{\ell}  (^tL)^K  \Psi (X ,
Y , Z , \theta ) | \ dY dZ d\theta . $$
Consequently:
$$h^{\ell /2}  |\nabla ^{\ell} \widetilde R_h ^{(N)} (F , G , X)|   \leq
C \sum _{\alpha + \beta \leq \ell + 2 K + 2N \atop \alpha \geq N ,
\beta \geq N}h^{ ( \alpha + \beta)/2} I_{\alpha \beta}(X , h) \leqno
(B.2) $$
$$ I_{\alpha \beta}(X , h)=   h^{ - 2n } \int
_{\R^{4n}\times [0, 1] } (1 - \theta)^{N-1}  \left ( 1 + {|X-Y|  +
|X-Z| \over \sqrt h} \right )^{-2K} | \nabla ^{\alpha} F (Y)| |
\nabla ^{\beta} G ( X + \theta (Z - X))| dY dZ d\theta.  \leqno
(B.3)$$
{\it End of the proof of  theorem 3.1.} We integrate the above
equality  (B.3) with respect to  $X$,  by making the change of
variables:
$$ X = (1 - \theta)^{-1} \ ( \widetilde X - \theta \widetilde Z )
\hskip 2cm Y = \widetilde Y \hskip 2cm Z = \widetilde Z. $$
We obtain :
$$ \Vert I_{\alpha \beta}(. , h) \Vert _{L^1(\R^{2n})}  \leq ... \hskip 7cm \ $$
$$ ... \leq C h^{-2n}
  \int _{\R^{6n}\times [0, 1] } (1 - \theta)^{ N - 2n -1} \left ( 1 + {|X-Y|  +
|X-Z| \over \sqrt h} \right )^{-2K} | \nabla ^{\alpha} F (Y)| \ |
\nabla ^{\beta} G (X ) |\ dX dY dY dZ d\theta .  $$
If  one  has  $ N \geq 2n+1$ and if one chooses  $K = 2n+1$, we
deduce, by using the Schur lemma, that:
$$ \Vert I_{\alpha \beta}(. , h) \Vert _{L^1(\R^{2n})}  \leq C
   \Vert  \nabla ^{\alpha} F
\Vert _{L^p (\R^{2n})} \ \Vert  \nabla ^{\beta} G \Vert _{L^{q
}(\R^{2n})}. $$
 Adding  these inequalities, we obtain:
$$h^{\ell /2}  |\nabla ^{\ell} R_h ^{(N, 1)} (F , G , X)|   \leq
C \sum _{\alpha + \beta \leq \ell + 2 K + 2N \atop \alpha \geq N ,
\beta \geq N}h^{ ( \alpha + \beta)/2} \Vert  \nabla ^{\alpha} F
\Vert _{L^p (\R^{2n})} \ \Vert  \nabla ^{\beta} G \Vert _{L^{q
}(\R^{2n})}. $$
By similarly proceeding  for
 $\widetilde R_h ^{(N)} (G , F , .)$, we arrive at the upper bound  (3.5)
 of  theorem 3.1.  Point (3.6)  is then  deduced by  proposition
 2.1.

 \bigskip

 {\it End of the proof of  theorem3.2.}  If  one chooses  $K =
 2n+1$, it follows from  (B.3) that
 $$\Vert I_{\alpha \beta}(. , h) \Vert _{L^{\infty}(\R^{2n})}  \leq C
   \Vert  \nabla ^{\alpha} F
\Vert _{L^{\infty} (\R^{2n})} \ \Vert  \nabla ^{\beta} G \Vert
_{L^{\infty }(\R^{2n})}. $$
By reporting in (B.2), then in  (B.1),  we obtain the  majorization
(3.7)  of theorem 3.2.

 \hfill \carre

\bigskip

\centerline{\bf  References}

\bigskip

[AN1] Z. Ammari, F. Nier, {\it Mean field limit for bosons and infinite
dimensional phase space analysis}, Ann. Henri Poincar\'e, {\bf 9}
(2008) 8, 1503-1574.

\medskip

[AN2] Z. Ammari, F. Nier, {\it Mean field limit for bosons and
propagation of Wigner measures}, J. Math. Phys, {\bf 50} (2009), 4,
042107.

\medskip

[BGGM] C. Bardos, F. Golse, A.D.  Gottlieb, N.J. Mauser, {\it Mean field
dynamics of fermions and the time-dependent Hartree-Fock equation},
J. Math. Pures Appl, 9, {\bf 82} (2003), 6, 665-683.

\medskip

[B] R. Beals, {\it Characterization of pseudo-differential operators and applications},
Duke Math. J.,{\bf 44} (1977),1, 45-57.

\medskip

[BR] A. Bouzouina, D. Robert, {\it Uniform semiclassical estimates for
the propagation of quantum observables,} Duke Math. J, {\bf 11}, 2
(2002), 223-252.

\medskip
[BdPF1]  A. Bove, G. da Prato, G. Fano, {\it An existence proof for the
Hartree-Fock time-dependent problem with  bounded two-body
interaction}, Comm. Math. Phys, {\bf 37} (1974), 183-191.

\medskip

[BdPF2]  A. Bove, G. da Prato, G. Fano, {\it On the Hartree-Fock
time-dependent Problem,} Comm. Math. Phys, {\bf 49} (1976) 25-33.

\medskip

[BH] W. Braun, K. Hepp, {\it The Vlasov dynamics and its fluctuations
in the $1/N$ limit of interacting classical particles, } Comm. Math.
Phys, {\bf 56} (1977), 101-120.

\medskip

[CV] A.-P.Calder{\`o}n, R. Vaillancourt, {\it A class of bounded
pseudo-differential operators}, Proc. Nat. Acad. Sci. U.S.A., {\bf
69} (1972), 1185-1187.

\medskip

[CR1] M. Combescure, D. Robert, {\it Semiclassical spreading of quantum
wave packets and applications near unstable fixed points of the
classical flow, } Asymptotic Anal, {\bf 14} (1997) 377-404.

\medskip

[CF] A. Cordoba, C. Fefferman, {\it Wave packets and Fourier Integral
operators,} Comm. in P.D.E, 3 ({\bf 11}) (1978), 979-1005.

\medskip

[DLERS] A. Domps, P. L'Eplattenier, P.G. Reinhard, E. Suraud, {\it The
Vlasov equation for Coulomb systems and the Husimi picture,} Annalen
der Physik, {\bf 6} (1997), 455-467.

\medskip

[ES] L. Erd\"os, B. Schlein, {\it Quantum dynamics with mean field
interactions: a new approach}, J. Stat. Phys, {\bf 134}, (2009), 3,
1171-1210.

\medskip

[F]  G.B. Folland, {\it Harmonic analysis in phase space}, Annals of
Mathematics Studies, {\bf 122}. Princeton University Press,
Princeton, NJ, 1989.

\medskip

[FKS] J. Fr\"ohlich, A. Knowles, S. Schwarz, {\it On the mean-field
limit of bosons with Coulomb two-body interaction}, Comm. Math.
Phys, {\bf 288} (2009), 3, 1023-1059.

\medskip

[GMP] S. Graffi, A. Martinez, M. Pulvirenti, {\it Mean field
approximation of quantum systems and classical limit}, Math. Models
Methods Appl. Sci, {\bf 13} (2003), 1, 59-73.

\medskip

[H]L. H\"ormander, {\it The analysis of} linear partial differential
operators. III. Pseudo-differential operators.  Springer, 1985.

\medskip

[L1] N. Lerner, {\it Some facts about the Wick calculus,} in {\it
Pseudodifferential Operators,} 135-174, Lecture Notes in Math. 1949,
Springer, Berlin 2008.

 \medskip
[P] F. Pezzotti, {\it Mean-field limit and semiclassical
approximation for quantum particle systems.} Rend. Math. Appl. {\bf
7}, 29, (2009), 3-4, 223-340.

\medskip

[PP] F. Pezzotti, M. Pulvirenti, {\it Mean-field limit and
semiclassical expansion of a quantum particle system.} Ann. Henri
Poincar\'e, {\bf 10} (2009), 1, 145-187.

\medskip
[RS] M. Reed, B. Simon, {\it Fourier alnalysis, self-adjointness},
Academic Press, San Diego, 1975.

\medskip

[R1] D. Robert, {\it Autour de l'approximation semiclassique}, Progress in Mathematics, {\bf 68}, Birkh�user Boston, Inc., Boston, MA, 1987.

\medskip
[R2] D. Robert, {\it  Propagation of coherent states in quantum
mechanics and applications,}  Partial differential equations and
applications, 181�252, S�min. Congr, 15, SMF, Paris, 2007.

\medskip

[R] C. Rondeaux, {\it Classes de Schatten d'op\'erateurs
pseudo-diff\'erentiels,} Ann. E.N.S, {\bf 17}, 1, (1984), 67-81.

\medskip
[S] R. Schubert, {\it Semiclassical localization in phase space}, PhD thesis, Ulm, 2001.

\medskip
[Sp] H. Spohn, {\it Kinetic equations from Hamiltonian dynamics}, Rev.
Mod. Phys, {\bf 52} (1980),  3, 569-615.

\medskip

[U1] A. Unterberger, {\it Oscillateur harmonique et op\'erateurs
pseudo-diff\'erentiels}, Ann. Inst. Fourier,XXXIX, 3, 1979.

\medskip

[U2] A. Unterberger, {\it  Les op\'erateurs m\'etadiff\'erentiels},
Lecure Notes in Physics {\bf 126}, 205-241 (1980).

\vskip 2cm

laurent.amour@univ-reims.fr

\bigskip

khodja@univ-reims.fr

\bigskip

jean.nourrigat@univ-reims.fr

\end